\documentclass[11pt]{article}
\usepackage{latexsym}
\usepackage{graphicx}
\usepackage{caption}
\usepackage{psfrag}
\usepackage{amsmath,amssymb,dsfont}

\usepackage[T1]{fontenc}
\usepackage[latin1]{inputenc}
\usepackage{color,pst-plot}

\usepackage{amsfonts,amstext,mathrsfs,upgreek}

\newcommand{\be}{\begin{equation}}
\newcommand{\ee}{\end{equation}}
\newcommand{\bea}{\begin{eqnarray}}
\newcommand{\eea}{\end{eqnarray}}

\newcommand{\nn}{\nonumber}

\def\e{{\,\rm e}}
\def\Li{{\,\rm Li}}

\input epsf

\begin{document}

\begin{titlepage}

\begin{flushright}
\end{flushright}
\vspace*{1.5cm}
\begin{center}
{\Large \bf Resummation of Threshold,  Low- and High-Energy \\[.4cm] Expansions for Heavy-Quark Correlators}\\[3.0cm]

{\bf David Greynat$^{\dag}$} and {\bf Santiago Peris$^{\dag \dag}$}\\[1cm]

$^{\dag}$IFAE, $^{\dag\dag}$Grup de F{\'\i}sica Te{\`o}rica, Dept. de Fisica \\ Univ. Aut{\`o}noma de Barcelona, 08193 Barcelona, Spain.\\[0.5cm]

\end{center}

\vspace*{1.0cm}

\begin{abstract}

With the help of the Mellin-Barnes transform, we show how to simultaneously resum the expansion of a heavy-quark correlator around $q^2=0$ (low-energy), $q^2 = 4 m^2$ (threshold, where $m$ is the quark mass) and $q^2\rightarrow-\infty$ (high-energy) in a systematic way. We exemplify the method for the perturbative vector correlator at $\mathcal{O}(\alpha_s^{2})$ and $\mathcal{O}(\alpha_s^{3})$. We show that the coefficients, $\Omega(n)$, of the Taylor expansion of the vacuum polarization function in terms of the conformal variable $\omega$ admit, for large $n$, an expansion in powers of $1/n$  (up to logarithms of $n$) that we can calculate exactly. This large-$n$ expansion has a sign-alternating component given by the logarithms of the OPE, and a fixed-sign component given by the logarithms of the threshold expansion in the external momentum $q^2$.

\end{abstract}

\end{titlepage}

\section{Introduction}

Two-point correlators of heavy quarks are very useful objects for extracting parameters of QCD such as, e.g., the quark masses and the strong coupling constant $\alpha_s$, both from experiment\cite{ Kuhn:2007vp,Chetyrkin:2009fv,Kuhn:2010vx} and also from the lattice\cite{Allison:2008xk}. However, even our perturbative knowledge of these functions is quite limited due to the complexity of the required diagrammatic calculation. For instance, in the case of the vector vacuum polarization function, only at $\mathcal{O}(\alpha_s^0)$ and $\mathcal{O}(\alpha_s^1)$ is the analytic result fully known \cite{Kallen:1955fb}. At $\mathcal{O}(\alpha_s^2)$, state-of-the-art techniques produce partial answers for this function only in the form of local expansions around three particular values of the external momentum, to wit $q^2=0$, $q^2\rightarrow-\infty$ and $q^2=4 m^2$ where $m$ is the heavy-quark mass. These correspond to the so-called low-energy expansion, the operator product expansion and the threshold expansion of the correlator.

A phenomenal effort has been devoted to the calculation of an ever-increasing number of coefficients in all of these expansions. For instance, for the $\mathcal{O}(\alpha_s^2)$ contribution to the vacuum polarization, no less than 30 terms are known from the low-energy expansion \cite{BCS,MMM}; although in most of all other situations this number is reduced to just a few \cite{Chetyrkin,CKS1,CKS2,CHKS,Czarnecki,Kiyo}. Ideally, of course,  one would like to be able to know the full function (including its imaginary part) and, consequently, a strategy has to be devised in order to reconstruct the function from the three expansions, even if this is only approximate.

Up to now, the method that has been employed in this reconstruction is that of Pad\'e Approximants \cite{PA}. Pad\'e Approximants are ratios of two polynomials whose coefficients are tuned so that they reproduce the Taylor expansion in $q^2$ around the origin of a given correlator, to the highest possible degree. However, even though the vacuum polarization is an analytic function and, therefore, admits a Taylor series around $q^2=0$, this is not so either at threshold or for $q^2\rightarrow -\infty$ due to the presence of non-analytic dependence in the form of square-roots and logarithms of the momentum. Consequently, all by themselves, Pad\'e Approximants cannot  approximate the correlator in the full complex plane, unless this nonanalytic dependence is somehow removed. In order to do this, a battery of judiciously chosen functions is used for subtracting the nonanalytic dependence on $q^2$ from the original correlator. Pad\'e Approximants are then   constructed for the subtracted correlator, rather than for the correlator itself. As pointed out in, e.g., ref. \cite{Kiyo,Hoang} this procedure has a certain degree of arbitrariness. This ambiguity in the procedure can potentially be the source of a systematic uncertainty, and it is important that this uncertainty be quantified.  Normally, this is attempted by varying among several of the choices made in the aforementioned Pad\'e construction. The results obtained with the Pad\'e method are certainly interesting but, as higher and higher precision is always in demand, we think it is also interesting that one could devise alternative methods which could confirm (or if need be, modify) these results.

Therefore, we think it would be desirable to have an alternative procedure which could reconstruct the full function from the three expansions in an unambiguous way, and with which we could have a better handle on the possible systematic uncertainties involved. In this paper we will present what we believe to be one such method. Furthermore,  unlike Pad\'{e}s which are essentially numerical, our reconstruction is totally analytic.

The main observation is that, in the conformal variable $\omega$ in which the $q^2$ cut complex plane is mapped onto a unit disk, the coefficients of the threshold expansion and the OPE determine the asymptotic behavior  of the coefficients of the low-energy expansion. This property is easiest to recognize in the language of Mellin transforms, thanks to a mathematical result  which is known as the Converse Mapping Theorem \cite{Flajolet:1995}.

An intuitive way to understand this is the following. A finite set of Taylor coefficients knows
nothing about the region of analyticity of the function. So, for example, even if you know a thousand Taylor coefficients, strictly speaking, you still do not know  the value for the radius of convergence. The reason is because the radius of convergence is a limit procedure and things can change completely from the coefficient a thousand and 1 onwards. Since the region of analyticity knows about the asymptotic tail of the Taylor coefficients, it is not strange that by giving precise information
on the properties of the function at the boundary of this region of analyticity (i.e. the threshold expansion and the OPE) one may gather information on this asymptotic tail of the Taylor coefficients.  How to make this statement useful beyond just words, is what we have tried to do in this work.

As it turns out, the asymptotic behavior of the low-energy coefficients sets in so quickly that already for the second or the third coefficient it is a good approximation. This results in an efficient way to compute them. Once all the Taylor coefficients are known (even if only approximately) one can resum them and reconstruct the full function. As we will see, this resummation expresses the result for the full function as a unique combination of polylogarithms (and derivatives thereof) plus a known polynomial. Although we have set up the problem within the context of a two-point correlator of a heavy quark, we think that the method we have developed can be of more general use, and could be applied also in other contexts where several expansions to a function with a given analytic structure are known. Mellin transforms have also proven very useful in other contexts in perturbative calculations \cite{Smirnov,Friot:2005cu,David,DavidSamuel}.

In the next section we will set up the necessary definitions and describe in detail our method and the relevant mathematical results it is based on. In section 3, we will test the method with the case of the $\mathcal{O}(\alpha_s^2)$ vacuum polarization function which, although not fully known analytically, is sufficiently known to allow interesting nontrivial checks. In particular, since 30 low-energy coefficients have been calculated up to now \cite{BCS,MMM}, we will be able to compare our prediction for them with  their exact value. In section 4, we will apply the method to the $\mathcal{O}(\alpha_s^3)$ vacuum polarization function, for which only a few terms in the low-energy, threshold expansion and OPE have been calculated \cite{BCS,MMM,Kiyo, Hoang,Chetyrkin:2006xg}. We will reconstruct the full function, make again predictions for the coefficients of its low-energy expansion, and compare them with those already present in the literature. The final section will be devoted to some conclusions and outlook. Finally, in the appendix, we compare our approximation to the exact result in the case of a known function, choosing the result at $\mathcal{O}(\alpha_s^0)$ as an illustrative example.

\section{Description of the method}

Let $\Pi(z)$ be a function of the complex variable $z$ with the following properties:

\begin{itemize}

\item It is analytic in the disc $|z|<1$ where, consequently, it admits a convergent expansion  when $z\to 0$ as

\begin{equation}
\Pi(z) \underset{|z|<1}{=} \sum_{n= 0}^\infty C(n)\; z^n \label{TaylorPi}\ ,
\end{equation}
which, from now on, we will call the ``Taylor'' expansion.

\item It has a cut along $1\leq \mathrm{Re}\ z< \infty$. In the limit $z\to 1$,  it can be expanded in the form

 \begin{equation}
\Pi(z) \underset{z\rightarrow1}{\sim} \sum_{ p,k} A(p,k)\; (1-z)^p \log^k(1-z) \ , \label{ThresholdPi}
\end{equation}
where $k$ are integers but $p$ may be integers or half-integers. This expansion will be called the ``threshold'' expansion.

\item Finally, it admits an expansion for $z \to - \infty$  such as
\begin{equation}
\Pi(z) \underset{z\rightarrow-\infty}{\sim} \sum_{p,k} B(p,k)\; \frac{1}{z^p} \log^k(-4z)\;, \label{OPEPi}
\end{equation}
where $p$ and $k$ are both integers. We will refer to this expansion as the ``OPE''.

\end{itemize}

The problem we are trying to solve is how to reconstruct the function $\Pi(z) $ which matches onto the above three expansions in the corresponding regions, in an analytic and systematic way. That this is possible is already guaranteed  by the Taylor expansion (which is convergent) and the theorem on analytic continuation. However, an exact reconstruction requires complete knowledge of the infinite set of coefficients $C(n)$. Realistically speaking, this type of information is certainly not available. When only a finite number of $C(n)$'s is known, one clearly has no other option but to resort to the OPE and Threshold expansions to try to reconstruct the original function $\Pi(z) $. It seems reasonable to imagine that the more terms one knows, the better one will be able to determine this function. However, to the best of our knowledge, a concrete method to analytically  reconstruct $\Pi(z) $ is not available in the literature. Therefore, we would like to explain how one can do this reconstruction in an analytic and systematic way. This we will do with the help of the Mellin transform, using  the vector two-point correlator for a heavy quark as an example, and  illustrating how the method works in practice.

Let us recall the definition of  the vacuum polarization function $\Pi(q^2)$ through the correlator of two
electromagnetic currents $j^\mu(x)=\bar q(x)\gamma^\mu q(x)$, where $q(x)$ is a heavy quark,
\begin{align}
\label{pidef}
\left(g_{\mu\nu}q^2-q_\mu q_\nu\right)\, \Pi(q^2)
\, = \, \,
- \,i
\int d^4x\, \e^{iqx}\left\langle \,0\left|T\, j_\mu(x)j_\nu(0)\right|0\,
\right\rangle
\,,
\end{align}
and where $q^\mu$ is the external four-momentum. In QCD perturbation theory, $\Pi(q^2)$ may be decomposed to ${\mathcal O}(\alpha_s^4)$ as
\begin{align}
\label{Pi}
\Pi(q^{2}) \, = \,&\,
\Pi^{(0)}(q^{2})
\, + \,\left(\frac{\alpha_{s}}{\pi}\right)\,
\Pi^{(1)}(q^{2})+\left(\frac{\alpha_{s}}{\pi}\right)^{2}\,
\Pi^{(2)}(q^{2})\ + \left(\frac{\alpha_{s}}{\pi}\right)^{3}\,
\Pi^{(3)}(q^{2})
\, + {\mathcal O}(\alpha_s^4 )\, .
\end{align}
For definiteness, $\alpha_s$  denotes the strong coupling constant in the $\overline{\mathrm{MS}}$ scheme at the scale $\mu=m_{pole}$, but the precise definition  is not important for the discussion which follows. Equation (\ref{Pi}) will be understood in the on-shell normalization scheme  where a subtraction at zero momentum has been made in such a way as to guarantee that $\Pi(0)=0$. As it is well known, the vacuum polarization in Eq. (\ref{Pi}) satisfies a once subtracted dispersion relation, i.e.
\begin{equation}\label{disprel}
    \Pi(q^{2}) = q^2 \int_{0}^{\infty} \frac{dt}{t (t-q^2-i \varepsilon)}\ \frac{1}{\pi} \mbox{Im}\,\Pi(t+i \varepsilon)\ .
\end{equation}
When massless cuts are disregarded (as we will assume henceforth), the spectral function in (\ref{disprel}) starts at $t=4 m^2$, where $m$ is the heavy quark pole mass. In this case, the function $\Pi(q^{2})$ may be expanded for $q^2\to 0$ in powers of
\begin{equation}
z=\frac{q^2}{4 m^2}\ ,
\end{equation}
and it admits a Taylor expansion, exactly as in Eq. (\ref{TaylorPi}). The other two expansions, Eqs. (\ref{ThresholdPi},\ref{OPEPi}) coincide with the threshold expansion (for $q^2 \rightarrow 4 m^2$) and the OPE (when   $q^2\rightarrow -\infty$).

\begin{figure}
\begin{center}
\includegraphics[width=6in]{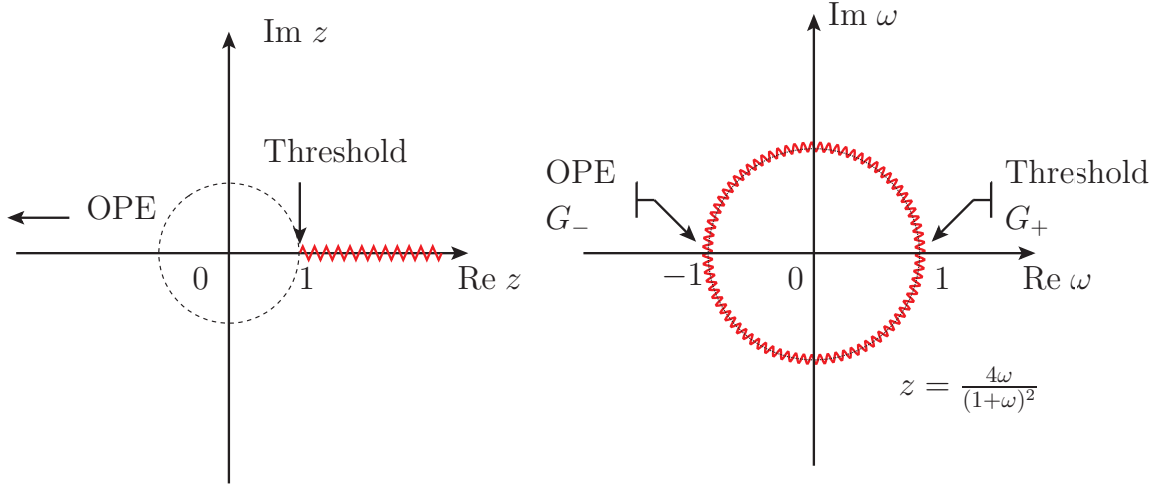}
\end{center}
\caption{Conformal mapping between $z$ and $\omega$, Eq. (\ref{conformal}). }
\label{omegaplan}
\end{figure}

In order to bring the three expansions (\ref{TaylorPi}-\ref{OPEPi}) to a common ground, it is very convenient to make the conformal transformation
\begin{equation}\label{conformal}
    z=\frac{4 \omega}{(1+\omega)^2}\qquad , \qquad \omega=\frac{1-\sqrt{1-z}}{1+\sqrt{1-z}}\ ,
\end{equation}
 which maps the cut $z$ plane into a unit disc in the $\omega$ plane, as we can see on figure \ref{omegaplan}. The cut $z\in [1, \infty[$ is transformed into the circle $|\omega| = 1$ and the points $z = 0$ into $\omega = 0$, $z =1$ into $\omega=1$, and the limit   $z \rightarrow +\infty \pm i \varepsilon$ into $\omega \rightarrow -1 \pm i \varepsilon$, and $z \rightarrow -\infty $ into $\omega \rightarrow -1$.

In terms of the new variable $\omega$, the vacuum polarization function $\widehat{\Pi}(\omega)$ also obeys the first and second expansions (\ref{TaylorPi}, \ref{ThresholdPi}), while the OPE becomes formally like the threshold expansion, except for a flip of sign in $\omega$. To wit:

\begin{itemize}

\item The Taylor expansion becomes

\begin{equation}
\widehat{\Pi}(\omega) \underset{|\omega|<1}{=} \sum_{n= 0}^\infty \Omega(n)\; \omega^n \label{TaylorPiw}\ ,
\end{equation}

\item whereas the OPE  ($\omega\rightarrow -1$) and threshold ($\omega\rightarrow +1$) expansion become

 \begin{equation}
\widehat{\Pi}(\omega) \underset{\omega\rightarrow \pm 1}{\sim}\sum_{\lambda, p} \Omega^{(\pm)}(\lambda, p)\; (1\mp \omega)^\lambda \log^p(1\mp \omega) \  . \label{ThresholdPiw}
\end{equation}

\end{itemize}

The two sets of coefficients, $ \Omega^{(+)}(\lambda, p)$ and $ \Omega^{(-)}(\lambda, p)$ are, in principle, unrelated as they correspond to expansions of the same function at two different points. However,  one can find the connection between the two sets of coefficients $C(n)$ in Eq. (\ref{TaylorPi}) and $\Omega(n)$ in Eq. (\ref{TaylorPiw}). They read

\begin{align}
\Omega(n) &= (-1)^n \sum_{p=1}^{n}  \frac{(-1)^p\;4^p\;\Gamma(n+p)}{\Gamma(2p)\Gamma(n+1-p)}\; C(p) \;,\label{Omega0=C}\\
 C(n)  &=\frac{\Gamma\left(n\right) \Gamma\left(\frac{1}{2}+n \right)}{\sqrt{\pi}}\;\sum_{p=1}^n \frac{ \Omega(p)\; p}{\Gamma\left(1+n-p\right) \Gamma\left(1+n+p\right)}\;.\label{C=Omega0}
\end{align}

As we will now see, the coefficients $\Omega^{(\pm)}(\lambda, p)$ in (\ref{ThresholdPiw}) determine the behavior of $\Omega(n)$ for large values of $n$. In order to see this, it is very convenient to redefine the conformal variable $\omega$ as
\begin{equation}\label{deft}
    \omega=\pm e^{-t}\ .
\end{equation}
The limit $t\rightarrow 0$ with the ``$+$'' sign corresponds to the threshold expansion, whereas the ``$-$'' sign corresponds to the OPE.  Then, one can make use of the integral representation
\begin{align}
\Gamma(s) &= \int_0^\infty\! dt \; t^{s-1} \; \e^{-t} & \rightleftharpoons  && \e^{-t} &= \int \limits_{c- i  \infty}^{c+ i  \infty} \! \frac{ds}{2i\pi} \; t^{-s} \; \Gamma(s)\; ,
\end{align}
where the integral on the righthand side term runs over a straight line in the $s$ complex plane, with $c$ any constant real number, $0< c< \infty$. The second identity allows one to express the vacuum polarization function (\ref{TaylorPiw}) as an inverse Mellin transform in the $t$ variable as
\begin{equation}
\widehat{\Pi}(\pm\e^{-t}) \underset{t>0}{=} \int \limits_{c- i \infty}^{c+ i  \infty} \! \frac{ds}{2i\pi} \; t^{-s} \; \Gamma(s) \; \sum_{n=1}^\infty \; (\pm 1)^n\; \Omega(n)\;  n^{-s} \label{MellinPit}\;.
\end{equation}
In this expression the integration path is a vertical line in the $s$ complex plane located at $Re(s)=c$, within the region of full analyticity of the integrand. This region is usually called ``the fundamental strip''.

The usefulness of the expression (\ref{MellinPit}) for our purposes is due to a relatively recent mathematical theorem, known as the \emph{Converse Mapping Theorem} \cite{Flajolet:1995}, which assures that the $t\rightarrow 0$ behavior of the function $\widehat{\Pi}(\pm\e^{-t})$ is completely determined by the singularities in the $s$ complex plane of the function appearing in the integrand of (\ref{MellinPit}), i.e.
\begin{equation}\label{integrand}
    \Gamma(s)\ G_{\pm}(s)\qquad \mathrm{where}\qquad G_{\pm}(s)=\sum_{n=1}^\infty \; (\pm 1)^n\; \Omega(n)\;  n^{-s}\; .
\end{equation}
Furthermore, these singularities are just limited to isolated poles, i.e. the above combination $\Gamma(s)\ G_{\pm}(s)$ is a meromorphic function. In general, the Converse Mapping Theorem  establishes the following dictionary \cite{Flajolet:1995} between a given function $f(t)$ and its Mellin transform $\mathcal{M}[f](s) $:
\begin{align}
\mathcal{M}[f](s) &= \int_0^\infty\! dt \; t^{s-1} \; f(t) & \rightleftharpoons  && f(t) &= \int \limits_{c- i  \infty}^{c+ i  \infty} \! \frac{ds}{2i\pi} \; t^{-s} \; \mathcal{M}[f](s) \label{dict1}\\
\mathcal{M}[f](s) & \asymp \sum_{p,k} \frac{r_{p,k}}{(s+p)^k} & \rightleftharpoons  && f(t) & \underset{t\rightarrow0}{\sim} \sum_{p,k} \frac{(-1)^{k-1}}{(k-1)!}\; r_{p,k}\;  t^p \log^{k-1} t \;. \label{dict2}
\end{align}
In the expression above the symbol $\asymp $ means ``singular expansion'' \cite{Flajolet:1995}, i.e. the sum of all the negative powers of the Laurent expansions of the function around every pole \footnote{Sometimes this expansion is also called the ``principal part'' of a meromorphic function.}. In particular, we would like to call the reader's attention to  the result shown in Eq. (\ref{dict2}), which says that the expansion of the function at $t\rightarrow 0$  is given  by the pole $p$ (together with its multiplicity $k$), and the residue $r_{p,k}$, of its Mellin transform. This property will allow us, for instance, to compute the behavior of the coefficients $\Omega(n)$ in the expansion (\ref{TaylorPiw}) for $n$ large.

One must first realize that the position and residue of a pole of the $G_{\pm}(s)$ function is entirely given by the asymptotic behavior as $n\rightarrow \infty$ of the $\Omega(n)$ coefficients. This is very important since it means that the function $ \Omega(n)$ admits an expansion in powers of $1/n$ for $n$ large.

For instance, if
\begin{equation}\label{example}
\Omega(n)=\frac{1}{n^p} \log^k n\ ,
\end{equation}
one finds that
\begin{equation}\label{asympt}
    G_{+}(s) = (-1)^k\ \zeta^{(k)}(s+p) \qquad , \qquad G_{-}(s) = (-1)^{k+1}\  \eta^{(k)}(s+p)
\end{equation}
where $\zeta^{(k)}(s)$ is the $k$-th derivative of the Riemann $\zeta$ function, and $\eta^{(k)}(s)$ is the $k$-th derivative of the Dirichlet $\eta$ function,
\begin{equation}\label{riemann}
\zeta(s) = \sum_{n=1}^\infty n^{-s}\hspace{1cm} \text{and} \hspace{1cm} \eta(s) = - \sum_{n=1}^\infty (-1)^n \; n^{-s} = \left(1-2^{1-s}\right) \zeta(s)\;.
\end{equation}
Now,
$$\zeta^{(k)}(s)\asymp (-1)^{k} \frac{\Gamma(k+1)}{ (s-1)^{k+1}}\ , $$
 exhibiting the pole at $s=1$. So the only pole of $G_{+}(s)$  in Eq. (\ref{asympt}) is located at $s=1-p$. One immediately sees that the location of this pole is correlated with the power of $n$ in the large-$n$ behavior of the $\Omega(n)$ coefficient (\ref{example}), while the multiplicity of the pole is given by the power of the $\log n$. On the contrary, the Dirichlet $\eta$ function has no pole. Of course the results in $G_{+}(s)$ and $G_{-}(s)$ get interchanged if the $\Omega(n)$ coefficient has the same large-$n$ behavior but with an alternating sign.

Looking at the vacuum polarization function $\widehat{\Pi}$ in Eq. (\ref{MellinPit}), one sees that one goes from the threshold expansion to the OPE by the addition of the alternating sign in the definition of the function  $G_{\pm}(s)$. This means that both expansions will be simultaneously encoded in a $\Omega(n)$ coefficient admitting an expansion for large $n$ of the form
\begin{equation}
\Omega(n)\underset{n\rightarrow \infty}{\sim} \Omega^{AS}(n) = \sum_{p,k} \left[ \alpha_{p,k}  + (-1)^n \beta_{p,k}\right]\frac{\log^k n }{n^p}\; . \label{DefOmega}
\end{equation}
In fact, the coefficients $\alpha_{p,k}$ encode the information from the non-analytic terms of the threshold expansion,  and the $\beta_{p,k}$ the information from the non-analytic terms in the OPE, as we will see explicitly in the next sections.

To illustrate the procedure, we will first treat the case of the $\Pi^{(2)}(q^2)$ contribution to the vacuum polarization, Eq. (\ref{Pi}). Since no less than 30 terms are known in the Taylor expansion of this function (\ref{TaylorPi}), this information will be very helpful to check our results and our understanding of the systematic errors involved. Furthermore, we will approximately reconstruct the function in the complex plane, including its imaginary part, and will be able to compute the value of a constant in the threshold expansion of this function which, up to now, has not been possible to obtain by ordinary diagrammatic methods. After that, we will move to the more interesting (and also more difficult) case of $\Pi^{(3)}(q^2)$, where we will repeat the same analysis, except that now only 3 coefficients from its Taylor expansion (\ref{TaylorPi}) together with just a few from the OPE and threshold expansion, are currently known.

\section{The case of $\Pi^{(2)}(q^2)$ }

Let us begin by considering the function $\Pi^{(2)}$. The threshold expansion (\ref{ThresholdPi}) has been calculated in \cite{Czarnecki,Hoang} to be
\begin{multline}\label{Pi2threshold}
\Pi^{(2)}(z) \underset{z\rightarrow 1}{\sim} \frac{A(-\tfrac{1}{2},0)}{\sqrt{1-z}} + \bigg\{A(0,2)\ \log^2(1-z) + A(0,1) \ \log(1-z) + A(0,0) \bigg\}\\
+ \bigg\{ A(\tfrac{1}{2},1)\ \log (1-z) + A(\tfrac{1}{2},0)\bigg\}\ \sqrt{1-z}+ \ldots
\end{multline}
in terms of certain known coefficients which we denote here as $A(m,n)$. For instance, $A(0,0)$ is the constant called  $K^{(2)}$ in reference \cite{Hoang}. In terms of the conformal variable $\omega$ (\ref{conformal}), this expansion reads (\ref{ThresholdPiw})
\begin{align}\label{Pi2omegaplus}
\widehat{\Pi}^{(2)}(\omega) \underset{\omega\rightarrow 1}{\sim}&\frac{\Omega^{(+)}(-1,0)}{1-\omega} + \bigg\{ \Omega^{(+)}(0,2)\  \log^2(1-\omega) + \Omega^{(+)}(0,1)\ \log(1-\omega) + \Omega^{(+)}(0,0)\bigg\}\nn \\
&+\bigg\{\Omega^{(+)}(1,0)+ \Omega^{(+)}(1,1) \ \log(1-\omega) \bigg\}\big(1-\omega\big)+ \ldots  \\ \nn
\end{align}
Obviously the two expansions (\ref{Pi2threshold}) and (\ref{Pi2omegaplus}) are related, with
\begin{align}
\Omega^{(+)}(-1,0) &= -2\  A\left(-\tfrac{1}{2},0\right)\nn \\
\Omega^{(+)}(0,2) &= 4 \ A(0,2) \nn \\
\Omega^{(+)}(0,1) &= 2 \ A(0,1) - 8 \ A(0,2)\ \log 2 \nn \\
\Omega^{(+)}(0,0) &= A(0,0) + A(-\tfrac{1}{2},0)-2\ A(0,1)\ \log 2 + 4\ A(0,2)\ \log^2 2 \nn \\
\Omega^{(+)}(1,0)&=A(0,1)-\frac{1}{2}\ A(\tfrac{1}{2},0)-4\ A(0,2)\ \log2 +  A(\tfrac{1}{2},1) \log2 \nn \\
\Omega^{(+)}(1,1)&=4\ A(0,2)- A(\tfrac{1}{2},1)\\
&\  \vdots \nn
\end{align}
As we have discussed (\ref{deft}), it is convenient to re-express the expansion (\ref{Pi2omegaplus}) in the $t$ variable, (recall that $\omega= \e^{-t}$), as
\begin{multline}
\!\!\!\!\widehat{\Pi}^{(2)}(\omega= \e^{-t}) \underset{t\rightarrow 0}{\sim} \frac{\Omega^{(+)}(-1,0)}{t} + \bigg\{\Omega^{(+)}(0,2)\  \log^2 t + \Omega^{(+)}(0,1)\ \log t
+ \Omega^{(+)}(0,0) + \frac{1}{2}\Omega^{(+)}(-1,0)\bigg\}\\
+ \bigg\{\frac{1}{12}\Omega^{(+)}(-1,0)- \frac{1}{2} \Omega^{(+)}(0,1)+ \Omega^{(+)}(1,0)+\left( \Omega^{(+)}(1,1)-\Omega^{(+)}(0,2)\right) \ \log t \bigg\}\ t +\ldots\;.
\end{multline}
This means, through the \textit{Converse Mapping Theorem} (\ref{dict1},\ref{dict2}), that the Mellin transform is given by \begin{align}
&\mathcal{M}\left[\widehat{\Pi}^{(2)}(\e^{-t})\right](s) = \Gamma(s)\ G_+(s)\nn \\
&\asymp \frac{\Omega^{(+)}(-1,0)}{s-1}\ +\  \frac{2\ \Omega^{(+)}(0,2)}{s^3}\ -\   \frac{\Omega^{(+)}(0,1)}{s^2}\
+\ \frac{ \Omega^{(+)}(0,0) + \frac{1}{2}\ \Omega^{(+)}(-1,0)} {s} \nn \\
&\;\;\; +\ \frac{ \Omega^{(+)}(0,2) - \Omega^{(+)}(1,1) } {(s+1)^2} + \ \frac{ \frac{1}{12}\ \Omega^{(+)}(-1,0)-\frac{1}{2}\ \Omega^{(+)}(0,1)+\Omega^{(+)}(1,0) } {s+1}+ \ldots\; ,\label{mellinthreshold}
\end{align}
where the last expression corresponds to its singular expansion. As one can see, the residues in this singular expansion are given by the coefficients of the threshold expansion (\ref{Pi2omegaplus}) in a very specific way.

Turning now to the OPE (\ref{OPEPi}), we have \cite{Chetyrkin}
\begin{multline}
\Pi^{(2)}(z) \underset{z\rightarrow -\infty}{\sim} \bigg\{ B(0,2)\ \log^2(-4z) + B(0,1)\ \log(-4z)+B(0,0)\bigg\}\\
+\bigg\{B(-1,2)\ \log^2(-4z)+B(-1,1)\ \log(-4z)+B(-1,0)\bigg\} \ \frac{1}{z} \\
+ \bigg\{B(-2,3)\ \log^3(-4z) + B(-2,2)\ \log^2(-4z)+B(-2,1)\ \log(-4z)+B(-2,0)\bigg\}\frac{1}{z^2}+ \ldots
\end{multline}
where the constants $B(0,0), B(-1,0)$ and $B(-2,0)$ are called $H_0^{(2)}, H_1^{(2)}$ and $H_2^{(2)}$ in reference \cite{Hoang}. In fact, many more terms of the OPE are known \cite{CHKS}, but we stop at $\mathcal{O}(z^{-2})$ because it is enough for our illustrative  purposes.

In terms of the conformal variable $\omega$ (\ref{conformal}),  one has
\begin{multline}\label{opeomega}
\widehat{\Pi}^{(2)}(\omega) \underset{\omega\rightarrow -1}{\sim}\bigg\{ \Omega^{(-)}(0,2)\log^2(1+\omega)+\Omega^{(-)}(0,1)\ \log(1+\omega)+\Omega^{(-)}(0,0) \bigg\}\\
+\bigg\{ \Omega^{(-)}(1,0)+ \Omega^{(-)}(1,1)\ \log(1+\omega) \bigg\} \big(1+\omega\big)\\
+\bigg\{ \Omega^{(-)}(2,0)+ \Omega^{(-)}(2,1)\ \log(1+\omega) +\Omega^{(-)}(2,2)\ \log^2(1+\omega)\bigg\} \big(1+\omega\big)^2
+ \cdots\;,
\end{multline}
where, again, the two expansions are related through
\begin{align}
\Omega^{(-)}(0,2) &=4 \ B(0,2) \nn\\
\Omega^{(-)}(0,1) &=-2\ B(0,1) - 16\ B(0,2)\ \log 2 \nn \\
\Omega^{(-)}(0,0) &= B(0,0) + 4\ B(0,1)\ \log 2 + 16\ B(0,2)\ \log^2 2 \;
\end{align}
and similar expressions for $\Omega^{(-)}(1,0), \Omega^{(-)}(1,1) , \Omega^{(-)}(2,0),\Omega^{(-)}(2,1)$ and $\Omega^{(-)}(2,2)$ which, because of their length, we do not write out explicitly.

Finally, in terms of the $t$ variable (recall that for the OPE one has that $\omega= - \e^{-t}$), this expansion reads
\begin{multline}
\widehat{\Pi}^{(2)}(\omega= -\e^{-t}) \underset{t\rightarrow 0}{\sim} \bigg\{\Omega^{(-)}(0,2)\log^2 t +\Omega^{(-)}(0,1)\log t +\Omega^{(-)}(0,0)\bigg\}\\
+\bigg\{ \frac{1}{24} \Omega^{(-)}(0,1) + \frac{1}{4} \Omega^{(-)}(0,2) - \frac{1}{2} \Omega^{(-)}(1,0)- \frac{1}{2} \Omega^{(-)}(1,1)+ \Omega^{(-)}(2,0) \\
+\bigg[\frac{1}{12} \Omega^{(-)}(0,2)- \frac{1}{2} \Omega^{(-)}(1,1) + \Omega^{(-)}(2,1) \bigg]\ \log t+ \Omega^{(-)}(2,2)\ \log^2 t\bigg\}\ t^2
+ \cdots\; .
\end{multline}
We emphasize that only even powers of $t$ may appear in this expansion. Consequently,  the Mellin transform (\ref{dict1},\ref{dict2}) becomes
\begin{multline}\label{mellinope}
\mathcal{M}\left[\widehat{\Pi}^{(2)}(-\e^{-t})\right](s)= \Gamma(s)\ G_-(s)\\
\asymp  2\  \frac{\Omega^{(-)}(0,2)}{s^3} -  \frac{\Omega^{(-)}(0,1)}{s^2} +  \frac{\Omega^{(-)}(0,0)}{s}\\
+ 2\ \frac{\Omega^{(-)}(2,2)  }{(s+2)^3}+ \frac{1}{12}\ \frac{6\ \Omega^{(-)}(1,1)- 12\ \Omega^{(-)}(2,1)- \Omega^{(-)}(0,2) }{(s+2)^2}\\
+  \frac{1}{24}\ \frac{\Omega^{(-)}(0,1)+ 6\ \Omega^{(-)}(0,2)- 12\  \Omega^{(-)}(1,0) - 12\ \Omega^{(-)}(1,1)+ 24\ \Omega^{(-)}(2,0)}{s+2}+ \ldots\; ,
\end{multline}
where the last expression corresponds to its singular expansion. Notice how the poles are located only at negative even numbers (besides zero) in correspondence with the even powers in the $t$ expansion. Again, the residues of the singular expansion are given by the coefficients of the OPE in (\ref{opeomega}) in a very specific way.

Following the discussion presented in section 2, the two singular expansions (\ref{mellinthreshold}) and (\ref{mellinope}) can be obtained by realizing that the $\Omega(n)$ coefficients admit an expansion for large values of $n$ of the form
\begin{align}
\Omega^{AS}(n)&=\alpha_{0,0} + \bigg\{\alpha_{1,0} + \alpha_{1,1}\ \log n\bigg\} \frac{1}{n} + \alpha_{2,0} \frac{1}{n^2}+\mathcal{O}\left(\frac{1}{n^3} \log^{\ell_1} n\right) \nn \\
& + (-1)^n \left[\bigg\{\beta_{1,0} + \beta_{1,1} \log n \bigg\}\frac{1}{n} + \bigg\{\beta_{3,0}  + \beta_{3,1} \log n \bigg\}\frac{1}{n^3} \right. \nn\\
& \hspace{2cm} + \left.  \bigg\{\beta_{5,0} + \beta_{5,1} \log n + \beta_{5,2} \log^2 n \bigg\}\frac{1}{n^5} +  \mathcal{O}\left(\frac{1}{n^7} \log^{\ell_2}n\right)\right]\label{asympt2}\;,
\end{align}
where the powers of the logarithms $\log^{\ell_{1},\ell_{2}}n$ correspond to the terms in the OPE and threshold expansion which have not been yet calculated. With this expression for $\Omega^{AS}(n)$, it is immediate to compute the $G_{\pm}(s)$ functions introduced in (\ref{MellinPit},\ref{integrand}). First we split the sum as
\begin{equation}\label{gplusminus}
G_{\pm}(s) =\sum_{n=1}^{\infty}(\pm)^n\ \Omega(n)\ n^{-s}= \sum_{n=1}^{\infty}(\pm)^n\ \Omega^{AS}(n) \ n^{-s}+ \sum_{n=1}^{\infty}(\pm)^n \left[\Omega(n)-\Omega^{AS}(n) \right] \ n^{-s}\ .
\end{equation}
Then, since some of the $\Omega(n)$ coefficients are known exactly, say for $n=1,...,N^*$, one can actually approximate the second sum as
 \begin{equation}\label{gplusminus1}
G_{\pm}(s) =\sum_{n=1}^{\infty}(\pm)^n\ \Omega^{AS}(n) n^{-s}+ \sum_{n=1}^{N^*}(\pm)^n \left[\Omega(n)-\Omega^{AS}(n) \right] n^{-s} + \delta G_{\pm}(N^*,s) \ .
\end{equation}
where
\begin{equation}\label{Mellinerror}
 \delta G_{\pm}(N^*,s)=\frac{1}{\Gamma(s)} \int_0^\infty dt\ t^{s-1}\ \mathcal{E}(N^*, \pm \mathrm{e}^{-t})\ ,
\end{equation}
and
\begin{equation}\label{error}
    \mathcal{E}(N^*, \omega)= \sum_{n=N^*+1}^{\infty} \ \left[\Omega(n)-\Omega^{AS}(n) \right] \ \omega^n\ .
\end{equation}
In other words, up to the $\Gamma(s)$ factor, $ \delta G_{\pm}(N^*,s)$ in (\ref{Mellinerror}) is the Mellin transform of the "error" function\footnote{An alternative expression is, of course, $ \delta G_{\pm}(N^*,s)= \sum_{N^*+1}^{\infty}(\pm)^n \left[\Omega(n)-\Omega^{AS}(n) \right]\ n^{-s} $, as one can see in (\ref{gplusminus1}). The important point is that this expression has to be understood by analytic continuation in $s$.} $\mathcal{E}(N^*, \omega)$. The above expression for $\mathcal{E}(N^*, \omega)$ makes it clear that the error in the approximation will  naturally become smaller as $N^*$ gets larger, or as more terms in the expansion (\ref{asympt2}) for $\Omega^{AS}$ are included. This is true over the whole physical region in the complex plane  $|\omega|\leq 1$.

Using the expansion $\Omega^{AS}(n)$ in Eq. (\ref{asympt2}) one can readily compute $G_{+}$ as
\begin{multline}\label{gplus}
G_+(s)=\alpha_{0,0}\ \zeta(s) +\alpha_{1,0}\ \zeta(s+1) - \alpha_{1,1}\ \zeta'(s+1) + \alpha_{2,0}\ \zeta(s+2) \\
+ \beta_{1,0}\ \eta(s+1) - \beta_{1,1}\ \eta'(s+1) + \beta_{3,0}\ \eta(s+3)-\beta_{3,1}\ \eta'(s+3) \\
+ \beta_{5,0}\ \eta(s+5) - \beta_{5,1}\  \eta'(s+5) + \beta_{5,2}\ \eta''(s+5) \\
+ \sum_{n=1}^{N^*} \left[\Omega(n)-\Omega^{AS}(n) \right] n^{-s} + \delta G_{+}(N^*,s) \qquad , \qquad
\end{multline}
and $G_{-}$ as
\begin{multline}\label{gminus}
G_-(s) = \beta_{1,0}\ \zeta(s+1) - \beta_{1,1}\ \zeta'(s+1) + \beta_{3,0}\ \zeta(s+3)-\beta_{3,1}\ \zeta'(s+3) \\
+ \beta_{5,1}\, \zeta(s+5) - \beta_{5,1}\ \zeta'(s+5) + \beta_{5,2}\ \zeta''(s+5) \\
+ \alpha_{0,0}\ \eta(s)+\alpha_{1,0}\ \eta(s+1) - \alpha_{1,1}\ \eta'(s+1) + \alpha_{2,0}\ \eta(s+2)\\ +  \sum_{n=1}^{N^*} (-1)^n\ \left[\Omega(n)-\Omega^{AS}(n) \right] n^{-s} + \delta G_{-}(N^*,s) \qquad , \qquad
\end{multline}
in terms of the Riemann $\zeta$ function and the Dirichlet $\eta$ function (recall Eqs. (\ref{riemann})) and their corresponding first and second derivatives. Matching the two expressions (\ref{mellinthreshold}) and (\ref{mellinope}) to the result in Eqs. (\ref{gplus}) and (\ref{gminus}) one readily obtains the following  results for the coefficients $\alpha_{i,j}$  and $\beta_{i,j}$ in (\ref{asympt2}):

\begin{align}
\alpha_{0,0} &= 2\ A(-\tfrac{1}{2},0) \quad (\ \simeq  3.44514) \nn \\
\alpha_{1,0} &= - 2\ A(0,1)+ 8\gamma_E A(0,2)+ 8\ \log2\ A(0,2) \quad (\ \simeq -0.492936 )\nn \\
\alpha_{1,1}&=8\ A(0,2) \quad (\ = 2.25)\nn \\
\alpha_{2, 0}& = A(\tfrac{1}{2},1)\quad (\ \simeq 3.05433)\nn
\end{align}
\begin{align}\label{asympt2values}
\beta_{1, 0}&=2\ B(0,1)+8\  \gamma_E B(0,2)+ 16\ \log2\ B(0,2) \quad (\ \simeq 0.33723) \nn\\
\beta_{1, 1}&=8\ B(0,2)  \quad (\ \simeq 0.211083) \nn \\
\beta_{3, 0}&=- B(1,1)+ 6\ B(-1,2) - 4\gamma_E B(-1,2)- \frac{2}{3} \ B(0,2) - 8 \log2\ B(-1,2) \ \  (\ \simeq 0.183422) \nn \\
\beta_{3,1}&=-4\ B(-1,2)   \quad (\ \simeq -0.620598 )\nn\\
\beta_{5,0}&= 3 B(-2,1) - 25 B(-2,2) + 12 \gamma_E\ B(-2,2) + 105 \ B(-2,3) - 150\gamma_E\ B(-2,3) \nn\\
&\;\;+ 36 \gamma_E^2\ B(-2,3) -6\pi^2 \ B(-2,3) - B(-1,1) + \frac{31}{3}\ B(-1,2) - 4 \gamma_E\ B(-1,2) \nn\\
&\;\;+ \frac{1}{15}\ B(0,2) + 24 \log 2 \ B(-2,2) - 300 \log 2 \ B(-2,3) + 144 \gamma_E \log 2 \ B(-2,3) \nn\\
&\;\;- 8 \log 2 \ B(-1,2) + 144 \log^2 2 \ B(-2,3)  \; \; (\simeq -4.30753) \nn\\
\beta_{5,1}&=12\ B(-2,2) - 150\ B(-2,3) + 72\ \gamma_E\ B(-2,3) + 144 \log 2\ B(-2,3) - 4\ B(-1,3) \nn\\
&(\simeq -1.89016) \nonumber\\
\beta_{5,2}&=36\ B(-2,3) \; \; (\simeq 1.38684) \quad  ,
\end{align}
where (in parentheses) we also give their approximate numerical value (for 3 light flavors). We emphasize that these $\alpha$ and $\beta$ coefficients in Eq. (\ref{asympt2}) are determined by the logarithms of the threshold expansion (\ref{ThresholdPi}) and the OPE (\ref{OPEPi}) in a totally unambiguous way\footnote{The coefficient $\alpha_{0,0}$ is determined by the squared-root singularity in the threshold expansion (\ref{Pi2threshold}), and is the only exception to this rule. This is all in agreement with the Converse Mapping Theorem \cite{Flajolet:1995}.}

\begin{figure}[h]
\begin{center}
\includegraphics[width=.8\textwidth]{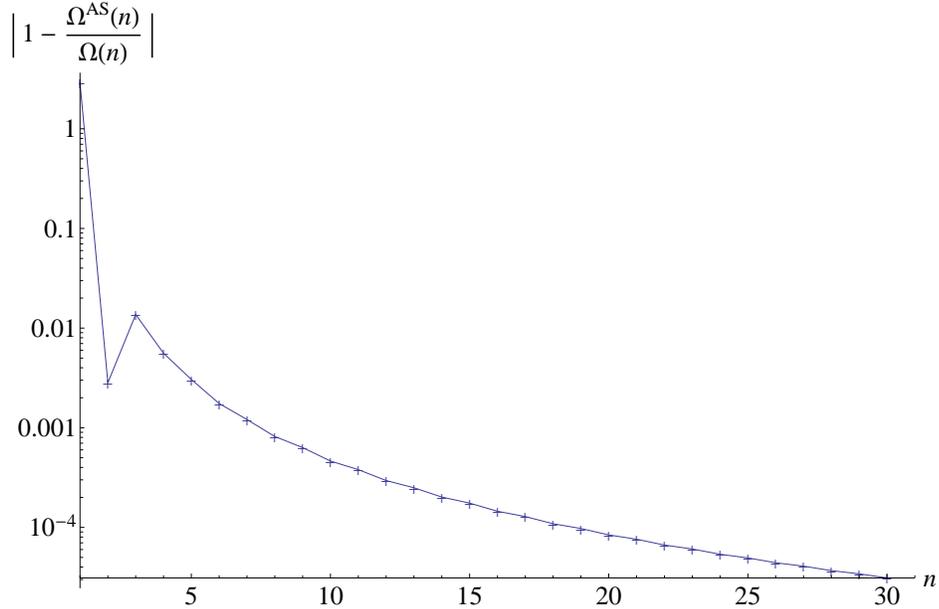}
\end{center}
\caption{Comparison of the $\Omega^{AS}(n)$ in Eqs. (\ref{asympt2}) and (\ref{asympt2values}) with the first 30 exact coefficients $\Omega(n)$.}
\label{Pi2OmegaAsOmega}
\end{figure}

Although the asymptotic expression (\ref{asympt2}) is, strictly speaking, only valid at large values of $n$, it is worth realizing that (\ref{asympt2}) sets in very quickly, and it becomes a very efficient approximation even for relatively low values of $n$. For instance, for $n\gtrsim 2-3$ the error is $\lesssim 1\%$, and gets better for larger values of $n$. This could be compared to the value predicted for $n=1$, which has a much bigger error of the order of $\sim 300\%$. To illustrate this fact, we plot in Fig. \ref{Pi2OmegaAsOmega} the exact value of the first 30 terms for $\Omega(n)$ (obtained from (\ref{Omega0=C}) with the exact results for the first 30 $C(n)$ from ref. \cite{CKS1,CKS2,BCS,MMM}) together with the result for the asymptotic expression (\ref{asympt2}, \ref{asympt2values}). This high efficiency of the asymptotic expression (\ref{asympt2}) will be turned into a method to predict the original Taylor coefficients $C(n)$ in (\ref{TaylorPi}) in the following.

 One can even get an idea of the $\mathcal{O}(n^{-3} \log^{\ell_1}n)$ error in the asymptotic expression (\ref{asympt2}) by performing a fit to the tail of the values for $C(n)$ for large $n$. Although the fit is very insensitive to the sign-alternating part of (\ref{asympt2}) and, essentially, yields no information on the term   $\mathcal{O}(n^{-7} \log^{\ell_2}n)$, the equal-sign term  $\mathcal{O}(n^{-3} \log^{\ell_1}n)$  is numerically rather well approximated by a term of the form $\sim + 0.5 \ n^{-3} \log^{1.5}n$. Taking a $100\%$ error on the value of the coefficient (but not on the sign) we estimate for the error:
 \begin{equation}\label{errorguess2}
    \left[\Omega(n) - \Omega^{AS}(n)\right]_{n>N^*} \cong  \left\{\begin{array}{c}
          +1 \\
          -0
        \end{array}\right\}\ \frac{\log^{1.5} n}{n^3} \; \pm \;\; (-1)^n \ \mathcal{O}\left(\frac{\log^{\ell_2} n}{n^7}\right)\; .
 \end{equation}
 Although the second term in the previous expression is essentially unknown, the large power  $n^7$ in the denominator makes it  essentially irrelevant for any reasonable value of $\ell_2$. This result will be useful for estimating the error of our calculations in what follows.

\subsection{Prediction for the constant $K^{(2)}$}

Given the result for $G_{+}(s)$  in (\ref{gplus}) it is straightforward to extract the residue at $s=0$ of the combination $\Gamma(s)G_{+}(s)$. According to Eqs. (\ref{MellinPit}-\ref{dict2}), this residue determines the constant $A(0,0)$ in Eq. (\ref{Pi2threshold}) (which is called $K^{(2)}$ in ref. \cite{Hoang}). This prediction is interesting because it has not been possible to obtain the value for this constant by ordinary diagrammatic methods up to now. The result yields
\begin{align}
&K^{(2)} = \nn\\
& -\frac{\alpha_{0,0}}{2} + \left(\frac{\pi^2}{12}+ \frac{\gamma_E^2}{2} + \gamma_1 \right)\alpha_{1,1}+  \frac{\pi^2}{6}\alpha_{2,0}- \zeta'(2)\ \alpha_{2,1} - \beta_{1,0}\log 2 + \left(-\frac{\log^2 2}{2} + \gamma_E \log 2\right)\beta_{1,1} \nn\\
&-\frac{3\zeta(3)}{4}\ \beta_{3,0} + \left(\frac{\zeta(3) \log 2}{4} + \frac{3\zeta'(3)}{4} \right)\beta_{3,1} - \frac{15}{16} \zeta(5)\ \beta_{5,0} + \left(\frac{\zeta(5) \log 2}{16} + \frac{15\zeta'(5)}{16}\right)\beta_{5,1} \nn\\
&+ \left(\frac{\zeta(5)\log^2 2 }{16} - \frac{\zeta'(5) \log 2}{8} - \frac{15\zeta''(5)}{16}\right)\beta_{5,2} +\sum_{n=1}^{N^*} \left[\Omega(n)-\Omega^{AS}(n) \right] + \mathcal{E}(N^*, 1)\; ,\label{K2}
\end{align}
where $\gamma_1\simeq -0.072816 $ is one of the Stieltjes's constants, appearing in the Laurent expansion of $\zeta(s)$ around $s=1$. In general, one has that
\begin{equation}\label{Stieltjes}
    \gamma_n=\lim_{x\rightarrow 1} \left [\zeta^{(n)}(x)- (-1)^n \frac{n!}{(x-1)^{n+1}}\right]\ .
\end{equation}
Although, in principle, the expression (\ref{K2}) depends on the unknown term $\mathcal{O}\left(n^{-7} \ \log^{\ell_2} n\right)$ in Eq. (\ref{errorguess2}) through $\mathcal{E}(N^*, 1)$ in Eq. (\ref{error}), the numerical result for $K^{(2)}$ is largely insensitive to this, for any reasonable value of the leading power $\ell_2$, due to the strong suppression of the $n^7$ factor in the denominator. Using Eq. (\ref{errorguess2}), our results in Eqs. (\ref{asympt2values})  and $N^*=30$ exact coefficients $C(n)$ from \cite{BCS,MMM}, neglecting the second term  $\mathcal{O}\left(n^{-7} \ \log^{\ell_2} n\right)$, one obtains
\begin{equation}\label{K2value}
    K^{(2)}= 3.783 \ ^{+ 0.004}_{-0.000} \ .
\end{equation}
This result is compared to two previous determinations, using Pad\'e techniques, in Table \ref{tableK2} .

\begin{table}
  \centering
\begin{tabular}{|c|c|}
  \hline
  Ref. & $K^{(2)}$ \\
  \hline
  \cite{Hoang} & $3.81\pm0.02$ \\
  \cite{Masjuan} & $3.71\pm 0.03 $\\
  \rule[-2ex]{0pt}{5ex} This work & $3.783 \ ^{+ 0.004}_{-0.000}$  \\
  \hline
\end{tabular}

\caption{Results for the threshold constant $K^{(2)}$. See text.}\label{tableK2}
\end{table}

We can now double check that the contribution from the  $\mathcal{O}\left(n^{-7} \ \log^{\ell_2} n\right)$ in Eq. (\ref{errorguess2}) is indeed small, for any reasonable value of the power $\ell_2$. In order to do this, let us note that
analogously to the previous case, the residue at $s=0$ of the combination  $\Gamma(s)G_{-}(s)$ also determines the value of the constant $B(0,0)$ (which was called $H_0^{(2)}$ in ref. \cite{Hoang}). The result reads

\begin{align}
&H^{(2)}_0 = \nn\\
&-\frac{\alpha_{0,0}}{2} - \log 2\, \alpha_{1,0} + \left(\gamma_E \log 2-\frac{\log^2  2}{2}\right)\,\alpha_{1,1}- \frac{\pi^2}{12}\, \alpha_{2,0} + \left(\frac{\pi^2 \log 2 }{12} + \frac{\zeta'(2)}{2}\right)\, \alpha_{2,1} \nn\\
&+\left(\frac{\pi^2}{12}+ \frac{\gamma_E^2}{2} + \gamma_1 \right)\beta_{1,1} + \zeta(3) \, \beta_{3,0} - \zeta'(3)\ \beta_{3,1} + \zeta(5)\ \beta_{5,0} -\zeta'(5)\ \beta_{5,1} + \zeta''(5)\ \beta_{5,2}\nn\\
& +\sum_{n=1}^{N^*} (-1)^n \left[\Omega(n)-\Omega^{AS}(n) \right] + \mathcal{E}(N^*, -1)\; .\label{H0}
\end{align}
However, unlike the case of $K^{(2)}$, the result for $H_0^{(2)}$ has been calculated\cite{CKS2} and, for 3 light flavors, it is given by
\begin{equation}\label{H02}
    H_0^{(2)}= -0.58570\quad .
\end{equation}
If we neglect the term $\mathcal{O}\left(n^{-7} \ \log^{\ell_2} n\right)$ in Eq. (\ref{errorguess2}), we can compute the value for the expression (\ref{H0}) with the help of our results in Eqs. (\ref{asympt2values}) and the $N^*=30$ exact coefficients $C(n)$ ($\Omega(n)$). The result obtained is $ H_0^{(2)}[Eq. (\ref{H0})]= -0.5856\ ^{+ 0.0001}_{-0.0000} \ $, in agreement with (\ref{H02}).

\subsection{Prediction for $C(n)$ coefficients}

\begin{figure}
\begin{center}
\includegraphics[width=0.9\textwidth]{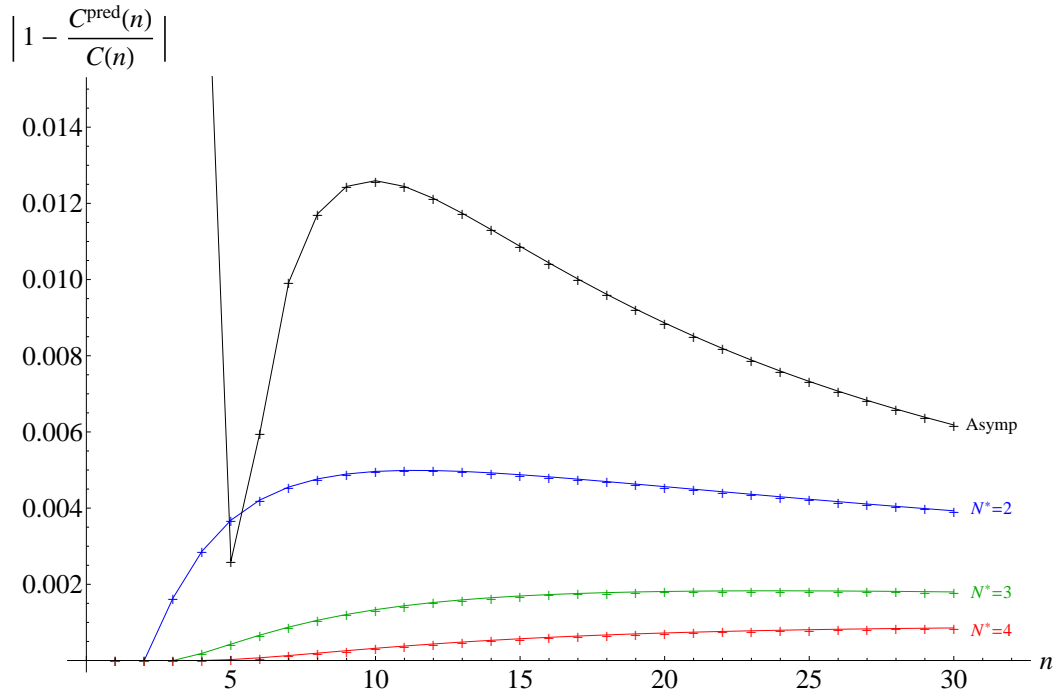}
\end{center}
\caption{Error of the approximation $C^{pred}(n)$ in Eq. (\ref{prediction2}) compared with the exact $C(n)$, for different values of $N^*=2$, $N^*=3$ and $N^*=4$, as explained in the text. Also shown is the direct asymptotic expression for $C(n)$ obtained in Ref. \cite{MMM} (labeled as ``Asymp'').}
\label{ErrorC2n}
\end{figure}

Expression (\ref{C=Omega0}) gives us the exact dictionary to translate the values of the coefficients $\Omega(n)$ into those for $C(n)$. Exact knowledge of a few $C(n)$ (let us say for $n=1,...,N^*$) together with the asymptotic expression $\Omega^{AS}(n)$, allows us then to make a \emph{prediction} for the rest of coefficients for $n>N^*$, as we will now describe.

Splitting the sum in  (\ref{C=Omega0}) between $1\leq p\leq N^*$ and $p>N^*$ and using the exact value of $\Omega(p)$ for the first part and the asymptotic expression and $\Omega^{AS}(p)$ (\ref{asympt2}) for the second part, one obtains the following prediction $C^{pred}(n)$:
\begin{multline}\label{prediction2}
C^{pred}(n) =\frac{\Gamma\left(n\right) \Gamma\left(\frac{1}{2}+n \right)}{\sqrt{\pi}}\;\left[\sum_{p=1}^{N^*} \frac{\Omega(p)\; p}{\Gamma\left(1+n-p\right) \Gamma\left(1+n+p\right)}\right.\\ \left.+ \sum_{p=N^*+1}^{n} \frac{\Omega^{AS}(p)\; p}{\Gamma\left(1+n-p\right) \Gamma\left(1+n+p\right)} \right]\;,
\end{multline}

The values obtained for $C^{pred}(n)$ can now be compared, e.g.,  with the result of the direct asymptotic expression for $C(n)$ obtained in ref. \cite{MMM} (black curve in Fig. \ref{ErrorC2n}). As one can see in this figure, already for $N^*=2$ (i.e. only assuming that the two coefficients $C(1)$ and $C(2)$ are exactly known), one achieves a better approximation than that obtained directly in terms of the coefficients $C(n)$ \footnote{Except for $n=5$, for which there seems to be a fortuitous cancelation.}, with errors which are  of the order of 6 per mil, at worst (blue curve). On this figure we also show the curves when $N^*=3$ (and $N^*=4$), i.e. when 3 (respectively 4) coefficients $C(n)$ are exactly known. In these cases, the errors are even smaller (green (respectively red) curves). This shows that Eq. (\ref{prediction2}) together with the asymptotic expression  (\ref{asympt2}, \ref{asympt2values}) is a good strategy to predict the value of the coefficients $C(n)$. This fact will be relevant in the case of  the function $\Pi^{(3)}$.

\subsection{Reconstruction of $\Pi^{(2)}(z)$}

\subsubsection{Analytic expression for $\widehat{\Pi}^{(2)}(\omega)$}
Starting from Eq. (\ref{TaylorPiw}), we can again rearrange the series by singling out the first $N^*$ coefficients, again assumed  to be exactly known, and write
\begin{equation}
\label{ApproxPi2v1}
\widehat{\Pi}^{(2)}(\omega) =\sum_{n=1} ^\infty \Omega^{AS}(n)\ \omega^n +\sum_{n=1} ^{N^*} \left[\Omega(n) - \Omega^{AS}(n)\right]\omega^n + \mathcal{E}(N^*,\omega)  \;,
\end{equation}
where
\begin{equation}\label{errorfunction}
\mathcal{E}(N^*,\omega) \doteq \sum_{n=N^*+1} ^\infty \left[\Omega(n) - \Omega^{AS}(n)\right]\omega^n \;,
\end{equation}
is the ``error function'' already defined in Eq. (\ref{error}). The first term in Eq. (\ref{ApproxPi2v1}) can be explicitly summed using the fact that\footnote{This function $\Li(s,\omega)$ is nothing else than the usual polylogarithm $\Li_s(\omega)$.  Here we write in this slightly unconventional way because we need to differentiate with respect to the variable $s$. We also use the notation $\Li^{(p)}(s,\omega)$ to signify $\frac{d^p}{ds^p}\Li(s,\omega)$.}
\begin{equation}
\sum_{n=1}^\infty \frac{\log^p n}{n^s}\ \omega^n = (-1)^p \Li^{(p)}(s,\omega)\; ,
\end{equation}
where the polylogarithm may be defined in the $\omega$ complex plane as
\begin{equation}\label{polylog}
    \Li(s,\omega)= \frac{-1}{\Gamma(s-1)}\ \int_0^1\frac{dx}{x}\ \log^{s-2}\left(\frac{1}{x}\right)\ \log(1-x\, \omega) \ .
\end{equation}

Consequently, we find the following expression for $\widehat{\Pi}^{(2)}(\omega)$,
\begin{align}\label{result2omega}
\widehat{\Pi}^{(2)}(\omega) &= \alpha_{0,0}\; \frac{\omega}{1-\omega} - \alpha_{1,0} \log(1-\omega) - \alpha_{1,1} \Li^{(1)}(1,\omega) + \alpha_{2,0}\Li(2,\omega)  \nonumber\\
& \hspace*{0.5cm}- \beta_{1,0}\log(1+\omega) - \beta_{1,1}\Li^{(1)}(1,-\omega) + \beta_{3,0}\Li(3,-\omega) - \beta_{3,1}\Li^{(1)}(3,-\omega) \nonumber\\
&\hspace*{0.5cm}+\beta_{5,0} \Li(5,-\omega) - \beta_{5,1} \Li^{(1)}(5,-\omega) + \beta_{5,2} \Li^{(2)}(5,-\omega) \nn\\
&\hspace*{0.5cm}+\sum_{n=1} ^{N^*} \left[\Omega(n) - \Omega^{AS}(n)\right]\omega^n + \mathcal{E}(N^*,\omega) \; ,
\end{align}
where the coefficients $\alpha_{i,j}$ and $\beta_{i,j}$  are given in (\ref{asympt2values}).

Neglecting the error function $\mathcal{E}(N^*,\omega) $ (an approximation which gets better as $N^*$ grows larger, and also as many more terms in the expansion (\ref{asympt2}) are known), our expression for $\widehat{\Pi}^{(2)}(\omega) $ is given by a unique combination of polylogarithms (and derivatives thereof) plus a known polynomial of degree $N^*$ in $\omega$. This expression constitutes a resummation of the Taylor expansion (\ref{TaylorPiw}) thanks to the information supplied by the threshold expansion  and the OPE, as seen in Eqs. (\ref{ThresholdPi}),(\ref{OPEPi}) and (\ref{asympt2values}). Notice that the knowledge of further terms in either expansion can only lead to the determination of further $\alpha$'s and $\beta$'s coefficients in $\Omega^{AS}(n)$, leaving unaltered the coefficients already determined.  Therefore, the approximation is systematic and can be improved upon by either knowing more $C(n)$ coefficients exactly (i.e. increasing $N^*$) or by knowing more terms from the threshold expansion or the OPE (i.e. by knowing more $\alpha$'s and $\beta$'s in $\Omega^{AS}(n)$). This should be clear from the fact that the expression for the error function $\mathcal{E}(N^*,\omega) $ in Eq. (\ref{errorfunction}) naturally becomes smaller in these two cases, for any $|\omega|\leq 1$ in the complex plane. In order to further illustrate this point, in the appendix we show the result of our resummation for the spectral function and compare it to the exact result, in the particular case of the function $\Pi^{(0)}(z)$ in Eq. (\ref{Pi}), where this result is fully known.

\begin{figure}[ht]
\begin{center}
\includegraphics[width=0.85\textwidth]{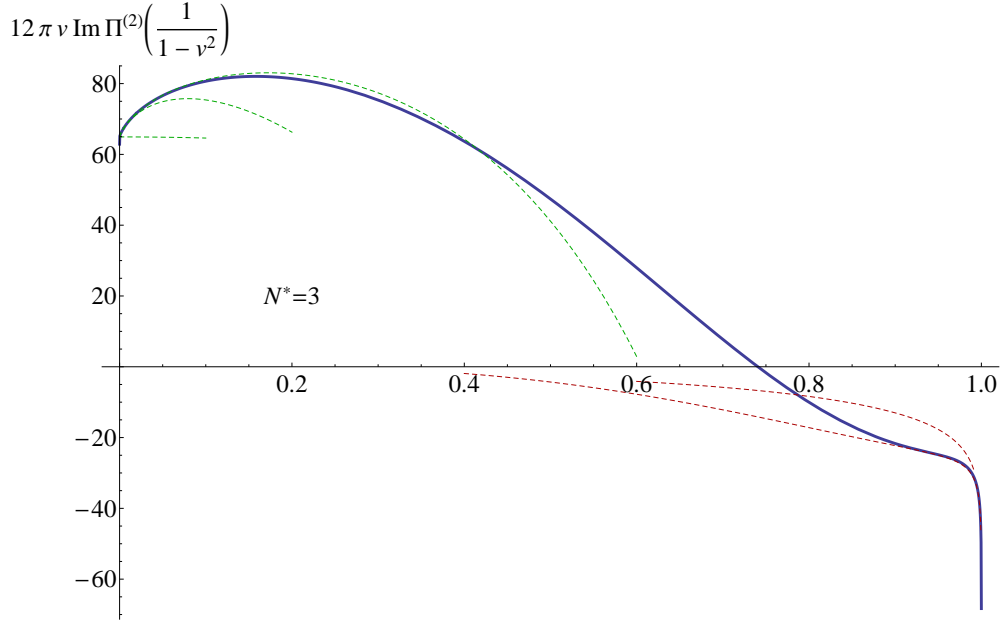}
\end{center}
\caption{Plot of the spectral function, ${\rm Im\,} \Pi^{(2)}\left(\frac{1}{1-v^2}\right)$, in terms of the quark velocity $v=\sqrt{1-1/z}$. The dashed curves correspond to consecutive approximations from the threshold expansion ($v\rightarrow 0$) and from the OPE ($v\rightarrow 1$).}
\label{Pi2ImPartN3}
\end{figure}

\subsubsection{Result for  Im$\Pi^{(2)}(z)$}

 Undoing the change of variables (\ref{conformal}), and using the expression for $ \widehat{\Pi}^{(2)}(\omega)$ in (\ref{result2omega}), one immediately obtains
\begin{equation}\label{piapprox2}
\Pi^{(2)}(z) = \widehat{\Pi}^{(2)}\left(\frac{1-\sqrt{1-z}}{1+\sqrt{1-z}}\right)\; ,
\end{equation}
as an representation for the function $\Pi^{(2)}(z) $ in the whole $z$ complex plane. In particular,
the analytic continuation of the polylogarithms and the polynomial in $\omega$ yield an approximation to its branch cut and the corresponding spectral function. In principle, we could easily give the explicitly analytic result for this spectral function from the expression (\ref{result2omega}, \ref{piapprox2}) but we refrain from doing so because the result is not particularly illuminating. Alternatively, we show in Fig. \ref{Pi2ImPartN3} the result for this spectral function as a function of the quark velocity $v=\sqrt{1- 1/z}$ for $N^*=3$, together with the corresponding approximations at threshold ($v\rightarrow 0$) and the OPE ($v\rightarrow 1$) (dotted curves). In this plot, the error band is estimated with the help  of the error function $\mathcal{E}(N^*=3, \omega)$ in (\ref{result2omega}) and the educated guess (\ref{errorguess2}).  Even for $N^*=3$ the errors are so small that they cannot be seen on the plot, becoming even smaller at larger values of $N^*$. This plot may be compared to the one in Fig. 3 of ref. \cite{Hoang}. One could of course use all the known $C(n)$ coefficients, i.e. take $N^*=30$ in our expressions. However, at the current level of precision, this is not yet necessary \cite{Hoang}.

\section{The case of $\Pi^{(3)}(q^2)$ }

We can now apply the same method exposed in the previous section for the $\Pi^{(2)}$ function, to treat the function $\Pi^{(3)}(q^2)$. For later convenience, let us first reproduce here the coefficients of the Taylor, OPE and threshold expansions listed in ref. \cite{Kiyo} for the vector vacuum polarization function at $\mathcal{O}(\alpha_s^3)$ (for 3 light flavors),
\begin{eqnarray}\label{expansion3}
  \Pi^{(3)}(z) \hspace{-.7cm}&&\underset{z\rightarrow 0}{=} 6.95646\ z + 7.24783\ z^2 + 7.31859\ z^3 + \mathcal{O}(z^4)\nn \\
   &&\underset{z\rightarrow 1}{=}\frac{2.63641}{1-z}-\frac{25.2331+7.75157\ \log(1-z)}{\sqrt{1-z}}+ K^{(3)}\nn \\
   &&\hspace{2cm} -11.0654\ \log(1-z)+1.42833\ \log^2(1-z)-0.421875\ \log^3(1-z)\nn \\
   &&\hspace{6cm} + \mathcal{O}\left(\sqrt{1-z}\ \log^\ell(1-z)\right)  \nn \\
   &&\underset{z\rightarrow -\infty}{=} H_0^{(3)}-0.0698778\ \log(-4z)+0.121085\ \log^2(-4z)-0.0366469\ \log^3(-4z)\nn \\
   &&+\frac{1}{z}\left\{H_1^{(3)}-3.75665\ \log(-4z)+2.11728\ \log^2(-4z)-0.318916\ \log^3(-4z) \right\}\nn \\
&&\hspace{-1.3cm}+   \frac{1}{z^2}\left\{H_2^{(3)}-5.13014\, \log(-4z)+0.31818\, \log^2(-4z)+0.401495\, \log^3(-4z) -0.0786514\, \log^4(-4z)\right\}\nn \\
&&\hspace{6cm}+ \mathcal{O}\left(z^{-3}\ \log^{\ell\, '}(-4z)\right)\qquad .
\end{eqnarray}

Using the coefficients of these expansions, a straightforward rerun of the method leads to the following result for the asymptotic expression of the corresponding coefficients $ \Omega^{AS}(n) $ in this case:

\begin{align}\label{OmegaAS3}
\Omega^{AS}(n) &= \alpha_{-1,0}\ n + \bigg \{ \alpha_{0,0} + \alpha_{0,1}\log n \bigg \}+\bigg\{\alpha_{1,0}+\alpha_{1,1}\log n + \alpha_{1,2}\log^2 n\bigg\}\frac{1}{n} + \mathcal{O}\left(\frac{1}{n^2} \log^{\ell_1}n\right)\nonumber\\
&\;\;+ (-1)^n \left[\bigg \{\beta_{1,0}+\beta_{1,1} \log n + \beta_{1,2}\log^2 n\bigg \}\frac{1}{n} + \bigg \{\beta_{3,0}+\beta_{3,1} \log n+\beta_{3,2}\log^2 n\bigg\}\frac{1}{n^3}\right.\nonumber\\
&\hspace{1.5cm}\left.  + \bigg \{\beta_{5,0}+\beta_{5,1}\log n + \beta_{5,2}\log^2 n + \beta_{5,3} \log^3 n \bigg\}\frac{1}{n^5}+ \mathcal{O}\left(\frac{1}{n^7} \log^{\ell_2}n\right)\right]\;,
\end{align}
with\footnote{Here we only quote the numerical results, for simplicity.}
\begin{equation}\label{values3}
\left\{
\begin{aligned}
\alpha_{-1,0}&\simeq 10.5456 \\
\alpha_{0,1} &\simeq 31.0063\\
\alpha_{0,0} &\simeq -11.0769 \\
\alpha_{1,0} &\simeq 36.3318 \\
\alpha_{1,1} &\simeq 37.1514\\
\alpha_{1,2} &\simeq 10.125
\end{aligned}\;,
\right.
\hspace{0.5cm}
\left\{
\begin{aligned}
\beta_{1, 0} &\simeq -0.181866\\
\beta_{1, 1} &\simeq -2.4852\\
\beta_{1, 2} &\simeq -0.879515\\
\beta_{3, 0} &\simeq -10.4385\\
\beta_{3, 1} &\simeq -4.7750\\
\beta_{3, 2} &\simeq 3.82702
\end{aligned}
\right.\;,
\hspace{0.5cm}
\left\{
\begin{aligned}
\beta_{5, 0} &\simeq -70.9277\\
\beta_{5, 1} &\simeq 56.3093 \\
\beta_{5, 2} &\simeq 20.9951 \\
\beta_{5, 3} &\simeq -7.55063
\end{aligned}
\right.\quad .
\end{equation}

\begin{figure}
\begin{center}
\includegraphics[width=5in]{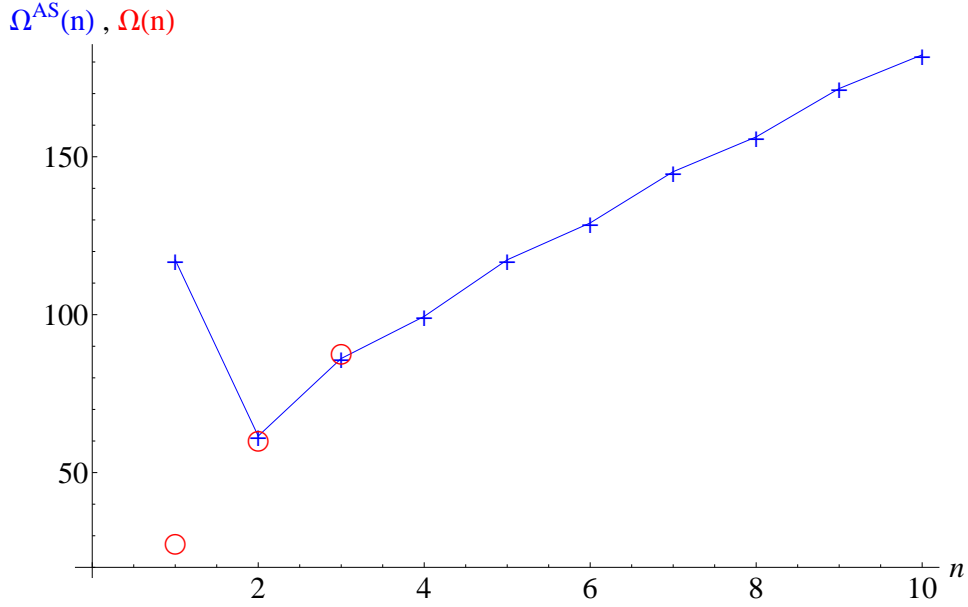}
\end{center}
\caption{Comparison between the asymptotic expression $\Omega^{AS}(n)$ in Eqs. (\ref{OmegaAS3}) and (\ref{values3})  (blue crosses) and the exact value for the first three coefficients $\Omega(n)$, $n=1,2,3$ (red circles). }\label{omega3plot}
\end{figure}

This expression for $\Omega^{AS}(n)$ can now be compared with the exact numerical result for $ \Omega(n)$ which can be obtained from the exact value of the first 3 coefficients $C(n)$, $n=1,2,3$, calculated in ref. \cite{Chetyrkin:2006xg,Boughezal:2006px,Maier:2008he,Maier:2009fz}, and our previous formula (\ref{Omega0=C}). The result is shown in Fig. \ref{omega3plot}. As one can see, the asymptotic form (\ref{OmegaAS3}, \ref{values3}) sets in very quickly, as it also happened in the previous case of the function $\Pi^{(2)}$ (see Fig. \ref{Pi2OmegaAsOmega}), reproducing  the exact values for $\Omega(2)$ and  $\Omega(3)$ within only a few per cent error. Notice that the error for $\Omega(1)$ would be much larger $\sim 300\%$, as it is clearly seen in Fig. \ref{omega3plot}. Based on this success for the ``prediction'' of $\Omega(2)$ and $\Omega(3)$, we can make a very rough guess for the error of the asymptotic expression (\ref{OmegaAS3})  in the form:
\begin{equation}\label{errorOmega3}
    \left[\Omega(n) - \Omega^{AS}(n)\right]_{n>N^*(\sim 2-3)} \cong \pm \ 15\ \frac{\log^{3} n}{n^2} \; \pm \;\; (-1)^n \ \mathcal{O}\left(\frac{\log^{\ell_2} n}{n^7}\right)\; .
 \end{equation}
At present we know nothing about the sign-alternating term $\sim \log^{\ell_2} n$ in the expression above although,  fortunately, one is protected by the high power in the denominator, i.e. $n^7$, for any reasonable value of the  ${\ell_2}$ power. This also happened in the case of the function $\Pi^{(2)}$ (see Eq. (\ref{errorguess2})). Therefore, the error coming from the first term in (\ref{errorOmega3}) and, moreover, its slow fall-off with $n$, i.e. $\sim n^{-2}$, will now be the main source of error in all our estimates. It is important to realize that the error (\ref{errorOmega3}) can only be reliably reduced once the terms of $\mathcal{O}(n^{-2} \log^{\ell_1}n)$  in Eq. (\ref{OmegaAS3}) are computed, which in turn requires the calculation of the terms $\mathcal{O}(\sqrt{1-z}\,\log^{\ell}(1-z))\ , (\ell\neq 0)$ in the threshold expansion (the second of the Eqs. (\ref{expansion3})). Regretfully, this information is not available at present. Alternatively, one could also compute more  $C(n)$ coefficients from the Taylor expansion, in order to increase the value of $N^*$. However, this second possibility does not seem so promising due to the rather slow $\sim n^{-2}$ fall-off with $n$ shown by the non sign-alternating part of the expression  in Eqs. (\ref{OmegaAS3}) and (\ref{errorOmega3}).

 At any rate, using the expression for $\Omega^{AS}(n)$ in (\ref{OmegaAS3},\ref{values3}), the result for the constant $K^{(3)}$ in Eq. (\ref{expansion3})\footnote{This constant is called $K_0^{(3),v}$ in Eq. (19) of ref. \cite{Kiyo}.}, akin to the constant $K^{(2)}$ in the section 3.1, now becomes
\begin{align}\label{K3}
&K^{(3)} = \nn\\
&-\frac{\alpha_{-1,0}}{12} - \frac{\alpha_{0,0}}{2} + \frac{\log 2\pi}{2}\, \alpha_{0,1} + \left(\frac{\pi^2}{12} + \frac{\gamma_E}{2} + \gamma_1 \right)\alpha_{1,1} + \left(\gamma_2-\frac{\gamma_E^3}{3}-\frac{\pi^2\gamma_E}{6}- \frac{2\zeta(3)}{3}\right)\alpha_{1,2}\nn\\
&+ \left(\frac{\gamma_E^4}{4} + \frac{\gamma_E^2\pi^2}{4}+ \frac{3\pi^4}{80} + \gamma_3+ 2\gamma_E \zeta(3)\right) \alpha_{1,3}
- \ \beta_{1,0} \log 2 + \left(-\frac{\log^2 2}{2} + \gamma_E \log 2\right)\beta_{1,1} \nn\\
&+ \left(-\frac{\log^3 2}{3} + 2\gamma_E\log^2 2 + 2\gamma_1 \log 2 \right) \beta_{1,2} + \left( \gamma_E \log^3 2 -\frac{\log^42}{4} + 3 \gamma_1 \log^2 2 + 3\gamma_2 \log 2 \right) \beta_{1,3} \nn\\
&- \frac{3\zeta(3)}{4}\ \beta_{3,0} + \left(\frac{\zeta(3)\log 2}{4} + \frac{3\zeta'(3)}{4} \right) \beta_{3,1} + \left(\frac{\zeta(3)\log^2 2}{4}+ \frac{\zeta'(3)\log 2}{2}+ \frac{3\zeta''(3)}{4} \right) \beta_{3,2} \nn\\
&+\left(\frac{\zeta(3) \log^3 2 }{4} - \frac{3\zeta'(3)\log^2 2}{4} + \frac{3\zeta''(3)\log 2}{4} + \frac{3\zeta^{(3)}(3)}{4} \right) \beta_{3,3} - \frac{15\zeta(5)}{16} \beta_{5,0}  \nn\\
&+ \left(\frac{3\zeta(5) \log 2}{48} + \frac{15\zeta'(5)}{16}\right) \beta_{5,1} + \left(\frac{\zeta(5)\log^2 2}{16} + \frac{\zeta'(5)\log 2}{8}- \frac{15\zeta''(5)}{16}\right) \beta_{5,2} \nn\\
&+\left(\frac{\zeta(5)\log^3 2}{16}-\frac{3\zeta'(5)\log^2 2}{16}+\frac{3\zeta''(5) \log 2}{16} + \frac{15\zeta^{(3)}(5)}{16} \right)\beta_{5,3} \nn\\
&+\left(\frac{\zeta(5)\log^4 2}{16} - \frac{\zeta'(5)\log^3 2}{4} + \frac{3\zeta''(5) \log^2 2}{8} - \frac{\zeta^{(3)}(5)\log 2 }{4} - \frac{14\zeta^{(4)}(5)}{16}\right) \beta_{5,4} \nn\\
&+\sum_{n=1} ^{N^*} \left[\Omega(n) - \Omega^{AS}(n)\right] + \mathcal{E}(N^*,1)\quad ,
\end{align}
where $\gamma_n$ are the Stieltjes's constants defined in Eq. (\ref{Stieltjes}). Using again the 3 exactly known coefficients $\Omega(n)$ ($n=1,2,3$) from \cite{Kiyo} (i.e. $N^*=3$) and our estimate for the error in the coefficients (\ref{errorOmega3}) to compute the error function $\mathcal{E}(N^*,1)$ defined in (\ref{error}), we obtain:
\begin{equation}\label{K3value}
    K^{(3)}= 18  \pm 87  \quad .
\end{equation}
Although our central value almost coincides  with that in ref. \cite{Kiyo}, our error is $\sim$ 8 times larger than that quoted in this reference and, in particular, cannot exclude a negative value for $  K^{(3)}$, as it was found in Ref. \cite{Hoang}. The situation is summarized in Table \ref{tableK3}.  It is now immediate to assess the impact of a reduction of the systematic error (\ref{errorOmega3}). For instance, if this error were reduced to $\sim\pm 15 \, n^{-3} \log^{4}n$ one would get an error in $K^{(3)}$, $\delta K^{(3)}\sim \pm 10$, whereas if the error could be pushed down to  $\sim\pm 15 \, n^{-3}$, the error would be $\delta K^{(3)}\sim \pm 0.6$; a dramatic improvement.

\begin{table}[h]
  \centering
\begin{tabular}{|c|c|}
  \hline
  Ref. & $K^{(3)}$ \\
  \hline
  \cite{Kiyo} & $17\pm 11 $\\
  \cite{Hoang} & $-10.09\pm 11$ \\
  \rule[-2ex]{0pt}{5ex} This work & $18 \pm 87$  \\
  \hline
\end{tabular}

\caption{Results for the threshold constant $K^{(3)}$. See text.}\label{tableK3}
\end{table}

As a cross-check of our method, we can also calculate the value of the first constant $ H_0^{(3)}$ appearing in the OPE \cite{Hoang}. We obtain
\begin{align}
&H_0^{(3)} = \nn\\
&-\frac{\alpha _{-1,0}}{4}-\frac{\alpha _{0,0}}{2}+\frac{\log \left(\frac{\pi }{2}\right)}{2}\  \alpha _{0,1} - \log 2\ \alpha _{1,0}+\left( -\frac{\log ^2 2}{2} + \gamma_E\log 2\right)  \alpha _{1,1}\nn \\
&+ \left(2 \gamma _1 \log 2 -\frac{\log ^3 2}{3} +  \gamma_E \log^2  2\right)  \alpha _{1,2}  + \left(3 \gamma_1 \log^2 2 +3 \gamma _2 \log 2 -\frac{\log ^4 2}{4} +\gamma_E  \log ^3 2 \right)\alpha _{1,3} \nn\\
&+\left(\frac{\pi^2}{12}+\frac{\gamma_E^2}{2} +\gamma_1\right) \beta _{1,1} + \left(-\frac{\pi^2\gamma_E}{6} -\frac{\gamma_E^3}{3} +\gamma _2-\frac{2\zeta (3)}{3}  \right)\beta _{1,2} \nn\\
&+ \left(\frac{3 \pi^4}{80} +\frac{\pi ^2\gamma_E^2}{4} +\frac{ \gamma_E ^4}{4} +\gamma _3 + 2 \gamma_E\zeta (3)\right)\beta _{1,3} +\zeta (3)\ \beta _{3,0} -\zeta '(3)\ \beta _{3,1} +\zeta ''(3)\  \beta _{3,2}   \nn \\
&-\zeta ^{(3)}(3)\ \beta _{3,3} +\zeta (5)\ \beta _{5,0} -  \zeta'(5)\ \beta _{5,1} +  \zeta''(5)\ \beta _{5,2} -\zeta ^{(3)}(5)\ \beta _{5,3}  + \zeta ^{(4)}(5)\ \beta _{5,4} \nn\\
&+\sum_{n=1} ^{N^*} \left[\Omega(n) - \Omega^{AS}(n)\right](-1)^n + \mathcal{E}(N^*,-1) \quad .
\end{align}
Using the values obtained in (\ref{values3}),  our estimate (\ref{errorOmega3}) and $N^*=3$, we obtain   $H_0^{(3)}=-7.1\pm 1.2$ while the exact value is $H_0^{(3)}=-6.1717$ \cite{Kiyo}. The reason why this estimate is much better than (\ref{K3value}) is because $K^{(3)}$ is more sensitive to the non sign-alternating component of the error (\ref{errorOmega3}), which is large. Further constants, such as $H_1^{(3)}$ and $H_2^{(3)}$ in (\ref{expansion3}),  are more sensitive to this component of the error, due to the extra powers of $n$ in the sums, resulting  in a larger total error and we refrain from giving their result.

Concerning our prediction for the coefficients $C^{pred}(n)$ , we can use the exact value for the first three $C(n)$, $n=1,2,3$, again from \cite{Chetyrkin:2006xg,Boughezal:2006px,Maier:2008he,Maier:2009fz}, and our result in  (\ref{prediction2}) with the expression for $\Omega^{AS}(n)$ from (\ref{OmegaAS3}) and the values (\ref{values3}) to obtain the results listed on Table \ref{coeff}.\footnote{There is factor $16\pi^2/3$ of difference between our normalization and that in Ref. \cite{Kiyo}.} We find agreement with the numbers quoted in \cite{Kiyo}, although our central values are systematically lower and the errors are larger.

\begin{table}
\begin{center}
\begin{tabular}{|c||c|c||c|c|}
\hline
$n$& $(16\pi^2/3)C^{pred}(n)$ for $N^*=3$ & Error&$ C(n)$ from \cite{Kiyo}& Error \\
 \hline
 1 & 366.174 & 0 & 366.175 & 0 \\
 2 & 381.510 & 0 & 381.509 & 0 \\
 3 & 385.235 & 0 & 385.233 & 0 \\
 4 & 382.7 & 0.5 & 383.073 & 0.011 \\
 5 & 378.0 & 1.2 & 378.688 & 0.032 \\
 6 & 372.5 & 1.8 & 373.536 & 0.061 \\
 7 & 367.0 & 2.3 & 368.23 & 0.09 \\
 8 & 361.5 & 2.7 & 363.03 & 0.13 \\
 9 & 356.4 & 3.1 & 358.06 & 0.17 \\
 10 & 351.6 & 3.4 & 353.35 & 0.2\\
\hline
\end{tabular}
\end{center}
\caption{Table of comparison between the $C^{pred}(n)$ from this work and the $C(n)$ in \cite{Kiyo}.}
\label{coeff}
\end{table}

Finally, we can also give the expression for $\widehat{\Pi}^{(3)}$  obtained by summing the coefficients (\ref{OmegaAS3},\ref{values3}), in complete analogy to what we did in Eq. (\ref{ApproxPi2v1}). The result is

\begin{align}\label{result3omega}
\widehat{\Pi}^{(3)}(\omega) & = \alpha_{\,-1,0} \; \frac{\omega}{(1-\omega)^2} + \alpha_{0,0} \; \frac{\omega}{1-\omega} - \alpha_{0,1} \Li^{(1)}(0,\omega) - \alpha_{1,0} \log(1-\omega) \nonumber\\
&- \alpha_{1,1}\Li^{(1)}(1,\omega)  + \alpha_{1,2} \Li^{(2)}(1,\omega) - \beta_{1,0} \log(1+\omega) - \beta_{1,1} \Li^{(1)}(1,-\omega) \nonumber\\
&+ \beta_{1,2}\Li^{(2)}(1,-\omega) + \beta_{3,0} \Li(3,-\omega) + \beta_{3,2} \Li^{(2)}(3,-\omega) + \beta_{5,0} \Li(5,-\omega) \nonumber\\
&- \beta_{5,1}\Li^{(1)}(5,-\omega) + \beta_{5,2}\Li^{(2)}(5,-\omega) - \beta_{5,3} \Li^{(3)}(5,-\omega) \nonumber\\
&+\sum_{n=1} ^{N^*} \left[\Omega(n) - \Omega^{AS}(n)\right]\omega^n + \mathcal{E}(N^*,\omega) \; .
\end{align}

\begin{figure}
\centering
\includegraphics[width=6in]{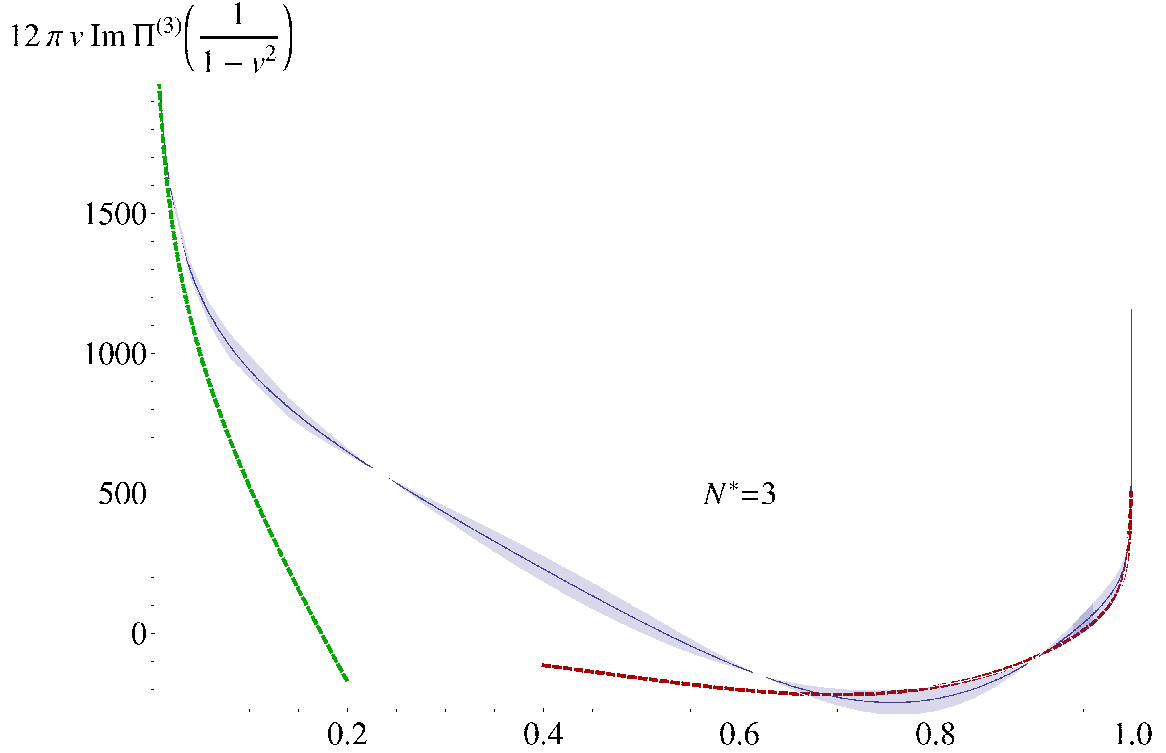}
\caption{Result for the spectral function of (\ref{result3omega}) in terms of the quark velocity, $v$, together with the threshold approximation ($v\rightarrow 0$) (in green) and the OPE ($v\rightarrow 1$) (in red).}\label{spectral3}
\end{figure}

Again, we plot in Fig. \ref{spectral3} the result for the spectral function which comes from the imaginary part of (\ref{result3omega}). This figure may be compared to the akin figures in ref. \cite{Hoang} (Fig. 5) and ref. \cite{Kiyo} (Fig. 2, first panel).

\section{Conclusions and Outlook}

Given an analytic function with a cut located at $1\leq z\leq \infty$,  we have presented a method  for reconstructing the full function from its Taylor expansion around the origin, i.e. $z\rightarrow 0$, Eq. (\ref{TaylorPi}), its threshold expansion for $z\rightarrow 1$, Eq.  (\ref{ThresholdPi}), and its OPE series for $z\rightarrow \infty$,  Eq. (\ref{OPEPi}). We have exemplified this method with the case of the perturbative vacuum polarization function of a  quark, but we think that it may be useful in other contexts as well.

The method uses the conformal variable $\omega$, Eq.  (\ref{conformal}), in terms of which the whole cut complex plane is mapped onto a disk of unit radius (with the cut running at the boundary, see Fig. \ref{omegaplan}), and consists in realizing that, in this variable, the Taylor coefficients $\Omega(n)$, Eq. (\ref{TaylorPiw}),  admit an asymptotic expansion in inverse powers of $n$, for large $n$, with a sign-alternating component and an equal-sign component. The coefficients of the equal-sign component, which we call $\alpha_{p,k}$  in Eq. (\ref{DefOmega}), are essentially determined by the logarithms of the threshold expansion, whereas the coefficients of the sign-alternating component, which we call $\beta_{p,k}$ in Eq. (\ref{DefOmega}), are determined by the logarithms of the OPE. Since logarithms are controlled by the Renormalization Group, it is not inconceivable that one may also link this asymptotic behavior in $n$ to the Renormalization Group in the future.

We have found that the precise dictionary, Eqs. (\ref{dict1},\ref{dict2}), is most easily established by means of the use of the Mellin transform and the Converse Mapping Theorem exposed in Ref. \cite{Flajolet:1995} and references therein. Once an asymptotic expression for the coefficients $\Omega^{AS}(n)$ is obtained, the power series can be explicitly summed using this asymptotic expression starting from a certain $N^*$ onwards. Below this $N^*$, the exact values for the coefficients must be used. The precision of the result of course depends on how precise the value of $\Omega^{AS}(n)$ is for $n=N^*$. Interestingly, we have found that the asymptotic series $\Omega^{AS}(n)$ sets in very quickly, giving quite accurate results already for $n\gtrsim 2,3$.

The resummed function in terms of the conformal variable $\omega$ can then be expressed as a combination of polylogarithms  and their derivatives plus a polynomial, with known coefficients (see, e.g. Eqs. (\ref{result2omega}), (\ref{result3omega}); and the Appendix). The method is analytic and it is possible to know how our results depend on our ignorance due to uncalculated terms in the Taylor, OPE and threshold expansion; which has given us a way to estimate the associated systematic error (\ref{error}).

We have made some numerical comparisons with the results obtained via the Pad\'{e} method \cite{PA,Kiyo,Hoang}, and our results agree with them, although our errors are usually larger. Nevertheless, we think it would be very interesting to be able to make a more thorough comparison including, in particular,  the result for the full spectral function. Furthermore, we plan to use our method to study other channels as well \cite{Future}.

Regarding the $\mathcal{O}(\alpha_s^3)$ vacuum polarization function, $\Pi^{(3)}$, we find that the calculation of all the terms $\mathcal{O}(\sqrt{1-z}\,\log^{\ell}(1-z))$ from the threshold expansion  for  $\ell\neq 0$, or at least the one with the highest power $\ell$, would result in a substantial reduction of the associated theoretical error in all the predictions for this function.

\vspace{1cm}

\textbf{Acknowledgements}

We thank P. Masjuan for his collaboration at an early stage of this work and A. Pineda for discussions.  We also thank S. Friot, M. Golterman, M. Jamin and E. de Rafael for comments on the manuscript.
This work has been supported by CICYT-FEDER-FPA2008-01430, SGR2005-00916, the Spanish Consolider-Ingenio 2010
Program CPAN (CSD2007-00042) and by the EU Contract No. MRTN-CT-2006-035482, ``FLAVIAnet''.

\vspace{1cm}

 \textbf{\Large APPENDIX }

 \vspace{.5cm}

In this Appendix we will use a known function to further illustrate our method. For this purpose, we will choose the $\mathcal{O}(\alpha_s^0)$ result, $\Pi^{(0)}(z)$,  which reads explicitly
$$\Pi^{(0)}(z)=\frac{3}{16 \pi^2}\left[\frac{20}{9}+\frac{4}{3\ z}-\frac{4(1-z)(1+2\ z)}{3\ z}\ G(z)\right]$$
where
$$G(z)=\frac{2\ u\  \log u}{u^2-1}$$
and
$$u=\frac{\sqrt{1-1/z}-1}{\sqrt{1-1/z}+1}\qquad .$$

Using the method we describe in the text,  the OPE and threshold expansions for this function determine that the asymptotic expansion of the associated coefficients $\Omega(n)$ in the Taylor expansion of $\widehat{\Pi}^{(0)}(\omega)$ in the conformal variable $\omega$ are given by
\begin{align}\label{asymptapp}
\Omega^{AS}(n) &= (-1)^n \left[-\frac{1}{2\pi^2}\ \frac{1}{n}+ \frac{9}{32\pi^2}\ \frac{1}{n^5} +...\right]\; .
\end{align}
Amusingly, there are only sign-alternating coefficients, i.e. $\beta_{i,j}$, but no $\alpha_{i,j}$ coefficient. This feature is in agreement with the fact that the threshold expansion of $\Pi^{(0)}(z)$ does not have any $\log(1-z)$ term, so all the terms in $\Omega^{AS}(n)$ are given by the $\log(-z)$ terms of the OPE.

\begin{figure}
\centering
\includegraphics[width=5in]{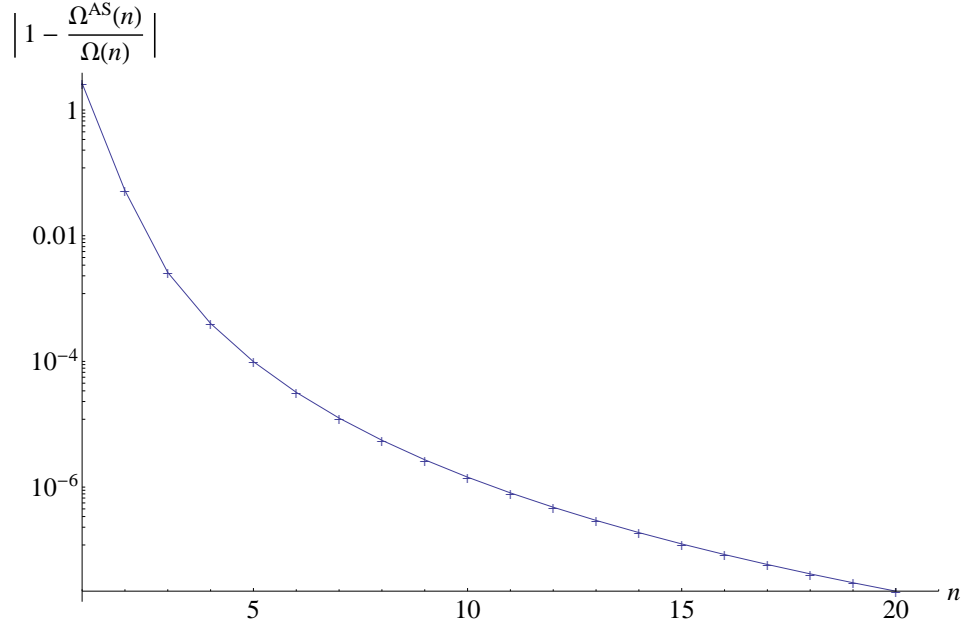}
\caption{Relative error between the asymptotic form $\Omega^{AS}(n)$  in Eq. (\ref{asymptapp}) and the exact values   for $\Omega(n)$ as a function of $n$.}\label{toymodelomega}
\end{figure}

As also happens for $\Pi^{(2)}(z)$ and $\Pi^{(3)}(z)$, the asymptotic expansion $\Omega^{AS}(n)$ sets in very quickly. In Fig. \ref{toymodelomega} we compare the exact values for the coefficients $\Omega(n)$ and this asymptotic expression. As one can see, already for $n=2$ the error is only a few percent, becoming smaller for larger $n$.

\begin{figure}
\centering
\includegraphics[width=5.5in]{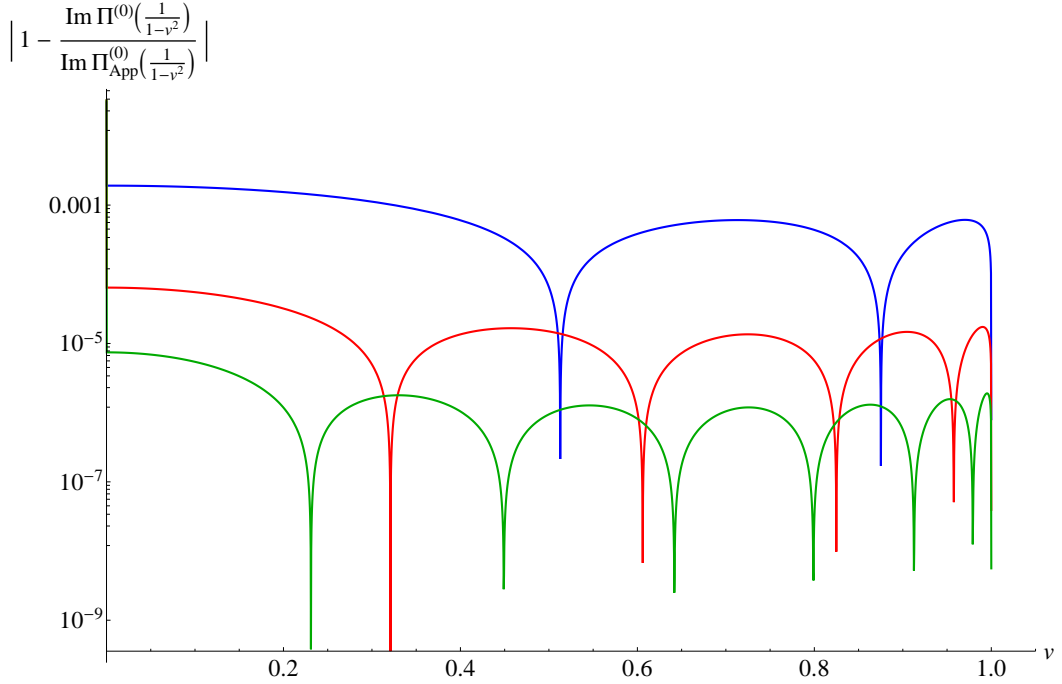}
\caption{Relative error in the spectral function for the approximation ${\Pi}^{(0)}(z)|_{App.} $ and the exact value  ${\Pi}^{(0)}(z) $ as a function of the quark velocity $v=\sqrt{1-1/z}$. The three curves shown correspond to $N^*=2$ (blue),  $N^*=4$ (red) and $N^*=6$ (green) }\label{toymodelspectrum}
\end{figure}

As to the approximation for the full function $\Pi^{(0)}(z)$, we obtain, in terms of the $\omega$ variable, the following expression:

\begin{eqnarray}
  \widehat{\Pi}^{(0)}(\omega)|_{App.} &= & \sum_{n=1}^{\infty}\Omega^{AS}(n)\ \omega^n+ \sum_{n=1}^{N^*}\left[ \Omega(n)-\Omega^{AS}(n)\right]\ \omega^n\nn \\
   &=& \frac{1}{2\pi^2}\log(1+\omega)+\frac{9}{32\pi^2}\Li(5,-\omega)+ \sum_{n=1}^{N^*}\left[ \Omega(n)-\Omega^{AS}(n)\right] \ \omega^n\qquad ,\nn
\end{eqnarray}
which, as usual, depends on how many coefficients $\Omega(n)$ we assume to be exactly known ($N^*$). In Fig. \ref{toymodelspectrum} we show how this expression approximates the exact function over the whole complex plane by comparing the spectral function obtained from this approximation to the exact spectral function of the function $\Pi^{(0)}(z)$, for different values of $N^*$.\footnote{Due to the dispersion relation (\ref{disprel}), it is clear that approximating the spectral function is equivalent to approximating the full function in the whole complex plane.}As we can see on this plot, even for $N^*=2$ (i.e. assuming only that the first two $\Omega(n)$ coefficients are exactly known) we reproduce the exact result with a relative error at the per mil level (or smaller), over the whole range in velocity $0\leq v\leq 1$. For larger values of $N^*$ the errors are, of course, even smaller.

\end{document}